\documentclass[twocolumn,showpacs,prb,aps,superscriptaddress]{revtex4}
\usepackage{amsmath}
\usepackage{graphicx}
\usepackage{dcolumn}
\usepackage{color}

\newcommand{\be}{\begin{equation}}
\newcommand{\ee}{\end{equation}}
\newcommand{\bea}{\begin{eqnarray}}
\newcommand{\eea}{\end{eqnarray}}
\newcommand{\bs}{\begin{split}}
\newcommand{\bse}{\begin{subequations}}
\newcommand{\ese}{\end{subequations}}

\begin{document}
\title{Superconductivity and Physical Properties of CaPd$_2$Ge$_2$ Single Crystals}
\author{V. K. Anand}
\altaffiliation{vivekkranand@gmail.com}
\affiliation {Ames Laboratory and Department of Physics and Astronomy, Iowa State University, Ames, Iowa 50011}
\affiliation{Helmholtz-Zentrum Berlin f\"{u}r Materialien und Energie, Hahn-Meitner Platz 1, D-14109 Berlin, Germany}
\author{H. Kim}
\author{M. A. Tanatar}
\author{R. Prozorov}
\author{D. C. Johnston}
\altaffiliation{johnston@ameslab.gov}
\affiliation {Ames Laboratory and Department of Physics and Astronomy, Iowa State University, Ames, Iowa 50011}

\date{\today}

\begin{abstract}

We present the superconducting and normal state properties of ${\rm CaPd_2Ge_2}$ single crystal investigated by magnetic susceptibility $\chi$, isothermal magnetization $M$, heat capacity $C_{\rm p}$, in-plane electrical resistivity $\rho$ and London penetration depth $\lambda$ versus temperature $T$ and magnetic field $H$ measurements. Bulk superconductivity is inferred from the $\rho (T)$  and $C_{\rm p}(T)$ data. The $\rho (T)$ data exhibit metallic behavior and undergoes a superconducting transition with $T_{\rm c\,onset} = 1.98$~K and zero resistivity state at $T_{\rm c\,0} = 1.67$~K\@. The $\chi(T)$ reveal the onset of superconductivity at 2.0~K\@. For $T>2.0$~K, the $\chi(T)$ and $M(H)$ are weakly anisotropic paramagnetic with $\chi_{ab} > \chi_{c}$. The $C_{\rm p}(T)$ confirm the bulk superconductivity below $T_{\rm c} = 1.69(3)$~K\@. The superconducting state electronic heat capacity is analyzed within the framework of a single-band $\alpha$-model of BCS superconductivity and various normal and superconducting state parameters are estimated. Within the $\alpha$-model, the $C_{\rm p}(T)$ data and the $ab$~plane $\lambda(T)$ data consistently indicate a moderately anisotropic $s$-wave gap with $\Delta(0)/k_{\rm B} T_{\rm c} \approx 1.6$, somewhat smaller than the BCS value of 1.764. The relationship of the heat capacity jump at $T_{\rm c}$ and the penetration depth measurement to the anisotropy in the $s$-wave gap is discussed.

\end{abstract}


\maketitle

\section{\label{Intro} Introduction}

With the advent of high-$T_{\rm c}$ superconductivity in FeAs-based compounds, these layered materials attracted much attention within the scientific community. \cite{Johnston2010, Canfield2010, Mandrus2010,Stewart2011}  Special interest in the doped $A$Fe$_2$As$_2$ ($A$ = Ca, Sr, Ba, Eu) class of materials arose because of their simple ${\rm ThCr_2Si_2}$-type body-centered tetragonal crystal structure (space group $I4/mmm$), the ease of growing large single crystals, and the similarity of their generic phase diagram showing the emergence of superconductivity upon suppression of antiferromagnetic spin density wave (SDW) transition with that of the high-$T_{\rm c}$ cuprates.\cite{Johnston2010, Canfield2010, Mandrus2010, Stewart2011, Johnston1997, Damascelli2003, Lee2006} The 122-type iron arsenide superconductors with the body-centered tetragonal ${\rm ThCr_2Si_2}$ structure consist of superconducting layers of ${\rm Fe_2As_2}$ stacked along the $c$~axis and FeAs layers appear to be crucial for the emergence of superconductivity.  However, the specific relationship of the FeAs layers and the mechanism of superconductivity in iron-arsenide superconductors is not well understood. Therefore the investigations of other 122-type materials without iron should be helpful in addressing such issues. The known iron-free 122-type Ca, Sr, or Ba-based pnictide superconductors include ${\rm BaNi_2As_2}$ ($T_{\rm c}=0.7$~K),\cite{Ronning2008} ${\rm SrNi_2As_2}$ ($T_{\rm c}=0.62$~K),\cite{Bauer2008} ${\rm SrIr_2As_2}$ ($T_{\rm c}=2.9$~K),\cite{Hirai2010} ${\rm SrPt_2As_2}$ ($T_{\rm c}=5.2$~K),\cite{Kudo2010} ${\rm SrNi_2P_2}$ ($T_{\rm c}=1.4$~K),\cite{Ronning2009} ${\rm BaNi_2P_2}$ ($T_{\rm c}=2.4$~K),\cite{Mine2008,Tomioka2009} ${\rm BaIr_2P_2}$ ($T_{\rm c}=2.1$~K), \cite{Berry2009} and ${\rm BaRh_2P_2}$ ($T_{\rm c}=1.0$~K),\cite{Berry2009} and ${\rm SrPt_2Sb_2}$ ($T_{\rm c}=2.1$~K).\cite{Imai2013} The compound ${\rm SrPd_2Ge_2}$ ($T_{\rm c}=3.0$~K) is a non-iron-based 122-type superconductor that is also free of pnictogen.\cite{Fujii2009,Kim2013} Non-bulk superconductivity was reported in ${\rm SrPt_2Ge_2}$ ($T_{\rm c}=10.2$~K) as arising from a high-temperature metastable ${\rm ThCr_2Si_2}$-type minority phase.\cite{Ku2013}

Continuing our investigations of the physical properties of the end-point compounds $AM_2$As$_2$ ($M$ = transition metal),\cite{Anand2012a,Anand2012b, Anand2013a} we recently discovered bulk superconductivity in the PdAs-based compounds ${\rm CaPd_2As_2}$ and ${\rm SrPd_2As_2}$ with $T_{\rm c}=1.27$~K and 0.92~K, respectively. They are both found to be conventional type-II $s$-wave superconductors.  Despite a very sharp jump in the electronic heat capacity $C_{\rm e}$ at $T_{\rm c}$ of ${\rm CaPd_2As_2}$, the value of $\Delta C_{\rm e}/\gamma_{\rm n} T_{\rm c}  =1.14(3)$ is significantly smaller than the BCS expected value of 1.43, where $\gamma_{\rm n}$ is the normal-state Sommerfeld electronic heat capacity coefficient.\cite{Anand2013a} The reduced value of $\Delta C_{\rm e}/\gamma_{\rm n} T_{\rm c}$ was accounted for by the $\alpha$-model of BCS superconductivity.\cite{Padamsee1973, Johnston2013}  The superconducting state $C_{\rm e}(T)$ data yielded a reduced $\alpha \equiv \Delta(0)/k_{\rm B}T_{\rm c} = 1.58(2)$ compared to the BCS value $\alpha_{\rm BCS} = 1.764$ which may be due to the presence of anisotropy in the $s$-wave superconducting order parameter (gap) and/or to multiple superconducting $s$-wave gaps.\cite{Johnston2013}

${\rm CaPd_2Ge_2}$ is reported to form in the ${\rm ThCr_2Si_2}$-type structure. \cite{Venturini1989} However the physical properties of this compound have not been reported previously. Here we report crystallographic and normal and superconducting-state properties of this compound.  We discovered that single-crystal ${\rm CaPd_2Ge_2}$ exhibits a very sharp bulk superconducting transition with a reduced value of $\Delta C_{\rm e}/\gamma_{\rm n} T_{\rm c}$, similar to ${\rm CaPd_2As_2}$.  Single crystals of ${\rm CaPd_2Ge_2}$ were grown by the high-temperature solution growth technique using PdGe as flux and were characterized by magnetic susceptibility $\chi$, isothermal magnetization $M$, heat capacity $C_{\rm p}$ and in-plane electrical resistivity $\rho$ measurements as a function of temperature $T$ and magnetic field $H$\@. For $T>2.0$~K the normal-state $\chi(T)$ is paramagnetic and weakly anisotropic with $\chi_{ab} > \chi_{c}$. A sharp jump in $C_{\rm p}$ at $T_{\rm c} = 1.69(3)$~K and zero resistivity below 1.67~K characterize ${\rm CaPd_2Ge_2}$ as a bulk superconductor. However $\Delta C_{\rm e}/\gamma_{\rm n} T_{\rm c}  =1.21(3)$ is found to be smaller than the BCS expected value of 1.426. In the superconducting state the electronic heat capacity data of ${\rm CaPd_2Ge_2}$ are well described by the $\alpha$-model\cite{Padamsee1973, Johnston2013} with a reduced $\alpha = 1.62(3)$ suggesting that the $s$-wave gap is anisotropic in momentum space\cite{Johnston2013} which is manifested as a reduced value of $\Delta C_{\rm e}/\gamma_{\rm n} T_{\rm c}$. Our penetration depth measurements indicate $s$-wave superconductivity with $\alpha = 1.56(2)$, consistent with the value of $\alpha$ inferred from the $C_{\rm p}(T)$ measurements.  Our measured and derived normal- and superconducting-state parameters characterize ${\rm CaPd_2Ge_2}$ as a conventional $s$-wave type-II superconductor in the dirty limit with a moderately anisotropic $s$-wave gap and/or with multiple $s$-wave gaps.

The experimental details are given in Sec.~\ref{ExpDetails} followed by a study of the crystallography in Sec.~\ref{Crystallography}. The superconducting and normal state properties are presented and analyzed in Secs.~\ref{Sec:CaPd2Ge2_Rho}--\ref{Sec:Ani_Gap}. A summary of the results and analyses is given in Sec.~\ref{Conclusion}.

\section{\label{ExpDetails} EXPERIMENTAL DETAILS}

Single crystals of ${\rm CaPd_2Ge_2}$ were grown by self-flux high-temperature solution growth using high purity Ca (Ames Lab), Pd (99.998\%, Alfa Aesar) and Ge (99.9999\%, Alfa Aesar) in a molar ratio of 1:4:4. The starting  materials were placed in an alumina crucible and sealed inside an evacuated quartz tube which was placed in a high-temperature furnace. The crystal growth was carried by heating to 1010~$^\circ$C at a rate of 80~$^\circ$C/h, holding for 40~h, then slow cooling over 100~h at a rate of 1.5~$^\circ$C/h down to 860~$^\circ$C, yielding several shiny plate-like crystals that were separated from the molten flux using a centrifuge.  Crystals of typical size $1.5 \times 1.2 \times 0.15$~mm$^3$ were obtained.

A JEOL scanning electron microscope (SEM) equipped with an energy-dispersive x-ray (EDX) analyzer was used to check the quality and chemical composition of the crystals. High-resolution SEM images revealed the crystals to be single phase and the EDX composition analysis showed the average composition of crystals to be consistent with the formula ${\rm CaPd_2Ge_2}$.  A Rigaku Geigerflex x-ray diffractometer was used to collect the powder x-ray diffraction (XRD) data using Cu~K$_\alpha$ radiation.  The crystal structure was determined by Rietveld refinement of the XRD data using the software {\tt FullProf}.\cite{Rodriguez1993}

A Quantum Design, Inc.\ superconducting quantum interference device magnetic properties measurement system (MPMS) was used for magnetic measurements. The anisotropic $M(T)$ and $M(H)$ data were collected by mounting a single crystal on a thin horizontal rotatable quartz rod using GE~7031 varnish and the contribution from the sample holder was subtracted to obtain the sample contribution.  The accuracy of the reported $\chi(T)$ and $M(H)$ data is $\approx10$\%.  A Quantum Design, Inc.\ physical properties measurement system (PPMS) was used for $C_{\rm p}(T)$ and in-plane $\rho (T)$ measurements. The $C_{\rm p}(T)$ was measured by the relaxation method using the heat capacity option of the PPMS and the $\rho (T)$ was measured by the standard four-probe ac technique using the ac transport option of the PPMS\@.  A $^3$He attachment to the PPMS allowed the $C_{\rm p}(T)$ and $\rho (T)$ measurements to be carried out down to a temperature of 0.45~K\@.

The temperature variation of the London penetration depth $\Delta\lambda(T)$ was measured using a tunnel diode resonator (TDR) operating in a dilution refrigerator at 20.2~MHz. The resonator consists of an $LC$ tank circuit with a single-layer copper coil with inductance $L_0\sim 1~\mu$H, a capacitor with capacitance $C = 100$~pF, and a tunnel diode that is biased to the region of negative differential resistance, thus compensating the losses in the circuit. This tank circuit develops self-oscillations at a resonant frequency $f_0=(2\pi \sqrt{L_0C})^{-1}$.  When a superconducting sample with magnetic susceptibility $\chi$ is inserted into the coil, the total inductance decreases and the (positive) resonant frequency shift $\Delta f(T) = f(T) - f_0$ is related to the temperature-dependent $\chi$ and hence $\lambda(T)$ by\cite{Prozorov2000, Prozorov2006}
\begin{equation}
\frac{\Delta f(T)}{G} =-4\pi \chi(T) \approx \left\{1 - \frac{\lambda(T)}{D}\tanh\left[ \frac{D}{\lambda(T)}\right]\right\},
\end{equation}
where $D$ is the effective sample dimension and $G$ is the sample- and coil-dependent calibration constant, $G \approx Vf_0/[2V_c(1-N)]$, where $V$ is the sample volume, $V_c$ is the coil volume and $N$ is the demagnetization factor of the sample. \cite{Prozorov2000}  To avoid uncertainties related to sample and coil dimensions as well as a poorly-defined demagnetization factor, the calibration constant $G$ is determined experimentally by matching the temperature dependence of the skin depth $d(T)$ obtained from the resonator response in the normal state to the measured resistivity $\rho(T)$ by using the relation $d(T)=(c/2\pi)\sqrt{\rho(T)/f}$, where $c$ is the speed of light in vacuum.\cite{Hardy1993}

A change in the London penetration depth with respect to its value at the lowest temperature $T_{\rm min}$ is measured according to $\Delta\lambda(T)=\lambda(T)-\lambda(T_{\rm min})$. In our experiment in a dilution refrigerator, $T_{\rm min} \approx 100$~mK, so $T_{\rm min}\ll T_c \approx 1.77$~K\@. The coil generates a very small ac magnetic field of about 20~mOe, which was oriented along the tetragonal $c$~axis of single crystals of CaPd$_2$Ge$_2$, so the reported London penetration depth corresponds to the penetration depth in the $ab$~plane.

\section{\label {Crystallography} Crystallography}

\begin{figure}
\includegraphics[width=3.3in]{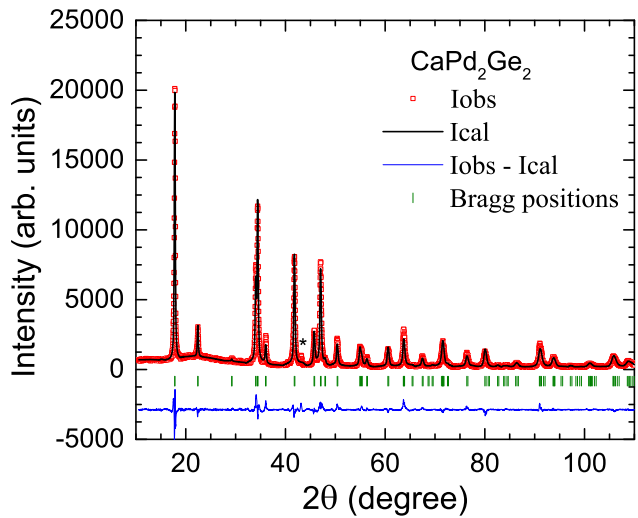}
\caption{ (Color online) Powder x-ray diffraction pattern of ${\rm CaPd_2Ge_2}$ recorded at room temperature using Cu~K$_\alpha$ radiation. The solid line through the experimental points is the Rietveld refinement profile calculated for the ThCr$_2$Si$_2$-type body-centered tetragonal structure (space group $I4/mmm$). The short vertical bars mark the fitted Bragg peak positions. The lowermost curve represents the difference between the experimental and calculated intensities. The unindexed peak marked with a star is a peak from residual PdGe flux on the surface of the ${\rm CaPd_2Ge_2}$ crystals prior to crushing them.}
\label{fig:CaPd2Ge2_XRD}
\end{figure}

\begin{table}
\caption{Crystallographic and Rietveld refinement parameters obtained from powder XRD data of crushed ${\rm CaPd_2Ge_2}$ crystals. The atomic coordinates of Ca, Pd and Ge atoms in space group $I4/mmm$ are (0,0,0), (0,1/2,1/4) and (0,0,$z_{\rm Ge}$), respectively.}
\label{tab:XRD1}
\begin{ruledtabular}
\begin{tabular}{llll}
Structure & ${\rm ThCr_2Si_2}$-type tetragonal  \\
Space group & $I4/mmm$ \\
\underline{Lattice parameters}\\
\hspace{0.8cm} $a$ (\AA)            		&  4.3216(1)  \\	
\hspace{0.8cm} $c$ (\AA)          			&  9.9675(3) \\
\hspace{0.8cm} $V_{\rm cell}$  (\AA$^3$) 	&  186.16(1)  \\
Ge $c$ axis coordinate $z_{\rm Ge}$   &  0.3745(2) \\
\underline{Refinement quality} \\
\hspace{0.8cm} $\chi^2$   & 6.88\\	
\hspace{0.8cm} $R_{\rm p}$ (\%)  & 7.57\\
\hspace{0.8cm} $R_{\rm wp}$ (\%) & 9.93\\
\end{tabular}
\end{ruledtabular}
\end{table}

The powder XRD data collected on crushed single crystals of CaPd$_2$Ge$_2$ at room temperature are shown in Fig.~\ref{fig:CaPd2Ge2_XRD} together with the Rietveld refinement profile for the ${\rm ThCr_2Si_2}$-type body-centered tetragonal structure (space group $I4/mmm$). The single-phase nature of crystals is evident from the refinement. The weak unindexed peak marked by a star is due to residual flux on the surface of the crystals prior to crushing them.  The crystallographic and refinement parameters are listed in Table~\ref{tab:XRD1}.  The lattice parameters $a$, $c$ and the $c$-axis position parameter $z_{\rm Ge}$ of the Ge atoms are in good agreement with the reported values.\cite{Venturini1989}  While refining the XRD data we kept the thermal parameters $B$ fixed to $B = 0$ and the occupancies of the atoms fixed to unity as there was no improvement in the quality of fit and the parameters $a$, $c$ and $z_{\rm Ge}$ were insensitive to small changes in $B$ and atomic site occupancies.  The $c/a$ ratio and the interlayer Ge--Ge distance $d_{\rm Ge-Ge} = (1-2z_{\rm Ge})c$ are 2.306(1) and 2.50~\AA, respectively.

\section{\label{Sec:CaPd2Ge2_Rho} Electrical Resistivity}

\begin{figure}
\includegraphics[width=3.3in]{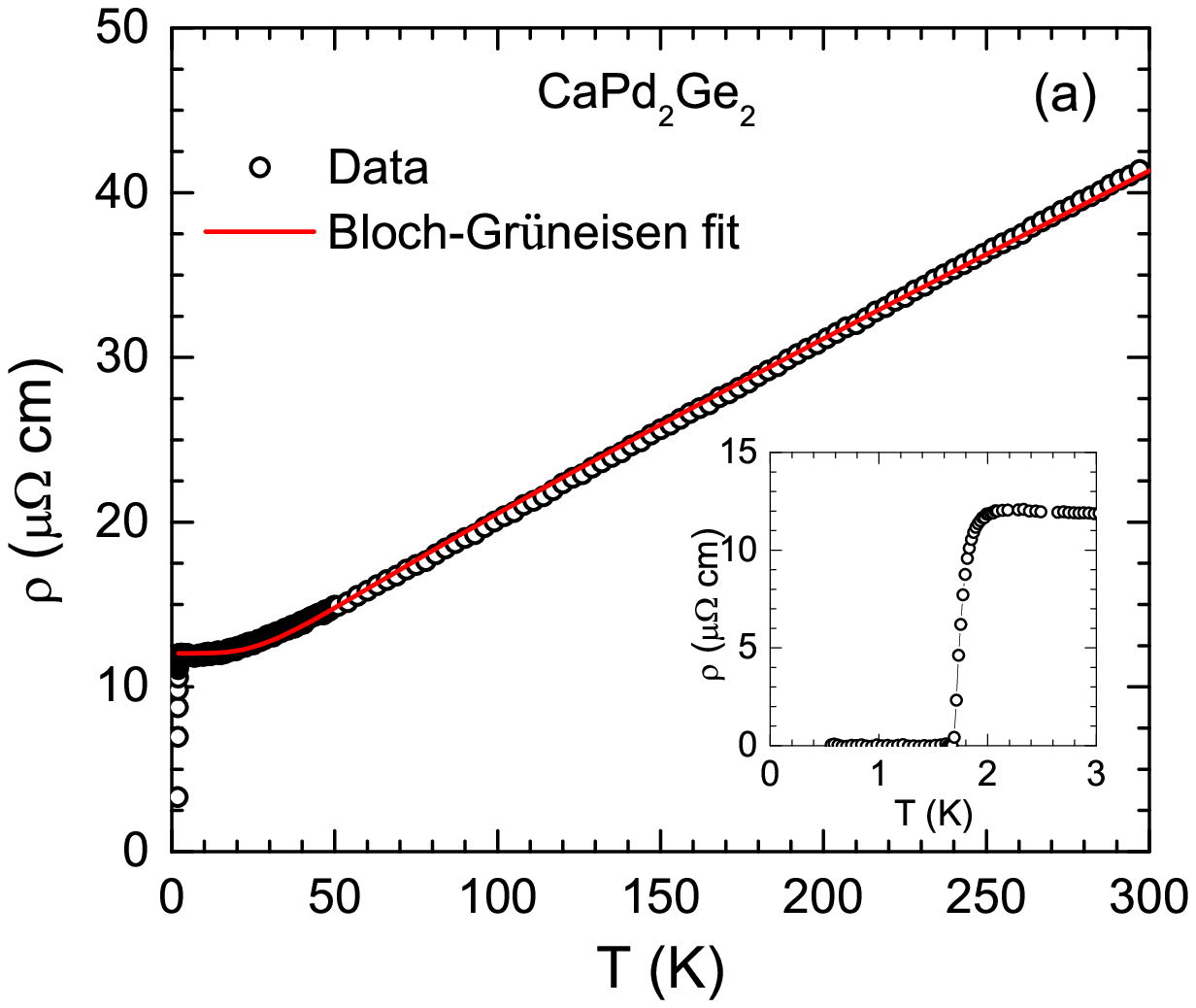}
\includegraphics[width=3.3in]{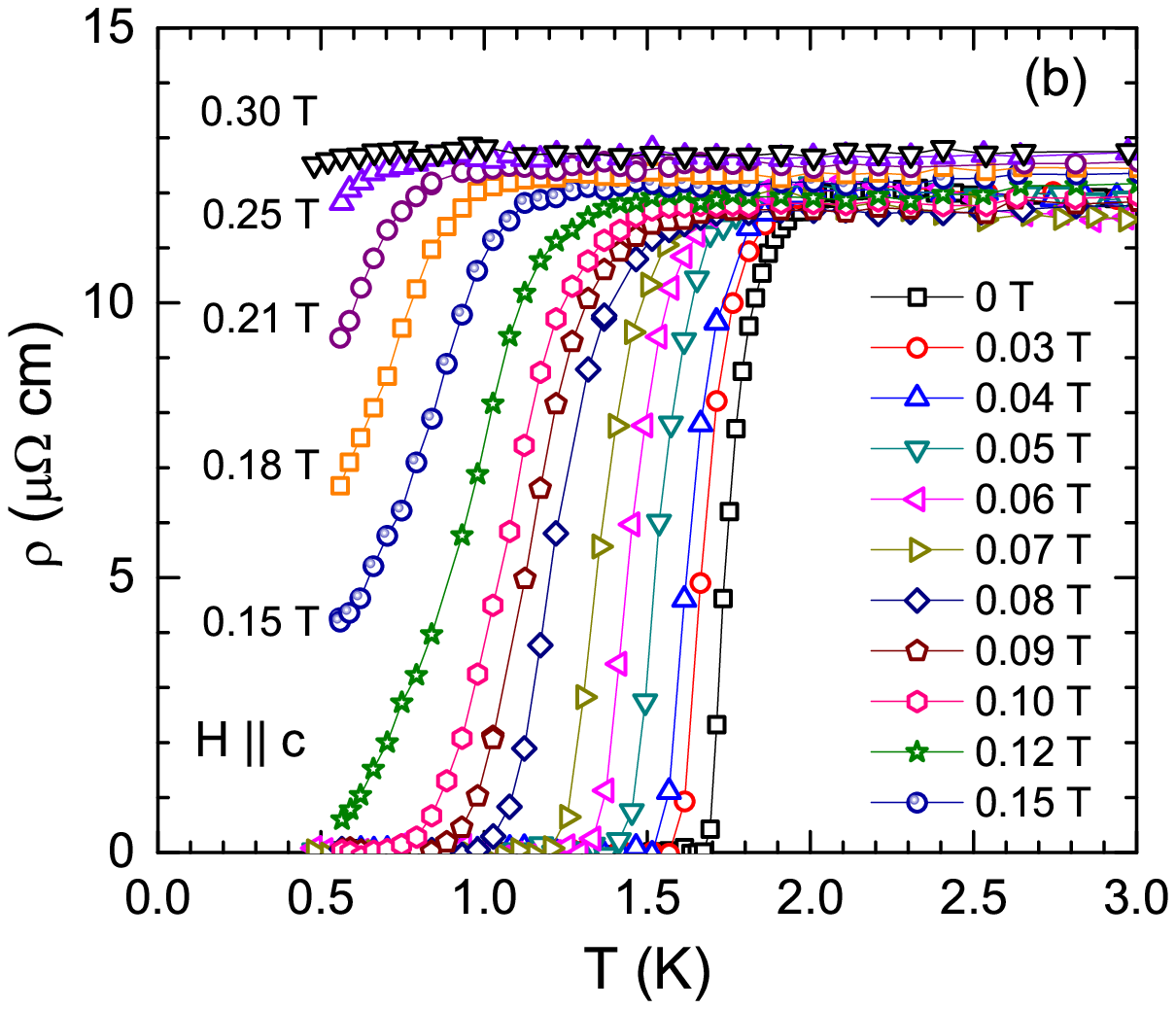}
\caption{(Color online) (a) In-plane electrical resistivity $\rho$ of a ${\rm CaPd_2Ge_2}$ single crystal as a function of temperature~$T$ for 1.8~K~$\leq T \leq$~300~K measured in applied magnetic field $H=0$. The red solid curve is a fit by the Bloch-Gr\"{u}neisen model. Inset: Expanded view of $\rho(T)$ below 3~K to show the superconducting transition. (b) The $\rho(T)$ of ${\rm CaPd_2Ge_2}$ for 0.45~K~$\leq T \leq$~3.0~K showing the superconducting transition for different values of $H$, as indicated, applied along the $c$~axis.}
\label{fig:CaPd2Ge2_Rho}
\end{figure}

The in-plane $\rho(T)$ data of CaPd$_2$Ge$_2$ measured at various fields in the temperature range 0.45~K~$\leq T \leq$~300~K are shown in Fig.~\ref{fig:CaPd2Ge2_Rho}. The $\rho(T)$ data shown in Fig.~\ref{fig:CaPd2Ge2_Rho}(a) reveals the metallic character of CaPd$_2$Ge$_2$. The $\rho$ decreases with decreasing $T$ and tends to be constant below 20~K until there is a rapid drop in $\rho$ to zero due to the onset of superconductivity. An expanded plot of the zero-field $\rho(T)$ data is shown in the inset of Fig.~\ref{fig:CaPd2Ge2_Rho}(a), which  exhibits a sharp superconducting transition with an onset at $T_{\rm c\,onset} = 1.98$~K and zero resistance  at $T_{\rm c\,0} = 1.67$~K\@. We define $T_{\rm c} = 1.71(5)$~K from the temperature of the peak of $d\rho/dT$\@. The $\rho(T)$ data in Fig.~\ref{fig:CaPd2Ge2_Rho}(b) at various $H$ show that $T_{\rm c} $ decreases with increasing $H$, where the transition is seen to broaden with increasing~$H$\@. While $T_{\rm c}$ is suppressed to a temperature below 0.5~K at $H \approx 0.12$~T, a $T_{\rm c\,onset}$ of 0.6~K is seen even at $H = 0.25$~T\@. In addition, an increase in the magnitude of the normal-state $\rho$ occurs with increasing $H$\@.

The normal-state residual resistivity at $T=2$~K and $H=0$ is $\rho_0 = 12~\mu \Omega$\,cm. The residual resistivity ratio  \mbox{RRR~$\equiv \rho(300\,{\rm K}) / \rho_0(2\,{\rm K}) = 3.5$}. The normal-state $\rho(T)$ data were fitted by the Bloch-Gr\"{u}neisen (BG) model of resistivity that describes the electrical resistivity due to scattering of conduction electrons by longitudinal acoustic lattice vibrations. \cite{Blatt1968} Our fit of $\rho(T)$ data by BG model is shown by the solid red curve in Fig.~\ref{fig:CaPd2Ge2_Rho}(a). Further details on the fitting procedure can be found in Refs.~\onlinecite{Anand2012a} and \onlinecite{Goetsch2012}. We obtained the resistively-determined Debye temperature $\Theta_{\rm{R}} = 171(3)$~K\@. The fit also yielded $\rho_0 = 12.03(3)~\mu \Omega$\,cm and $\rho(T = \Theta_{\rm{R}}) = 16.1(3)~\mu \Omega$\,cm.  The error bars on the two fitted resistivity values are statistical and do not take into account the $\sim10$\% uncertainty in the geometric factor that converts the measured resistance values into resistivity values.

\section{\label{Sec:CaPd2Ge2_M(H,T)} Normal-State Magnetization and Magnetic Susceptibility}

\begin{figure}
\includegraphics[width=3.3in]{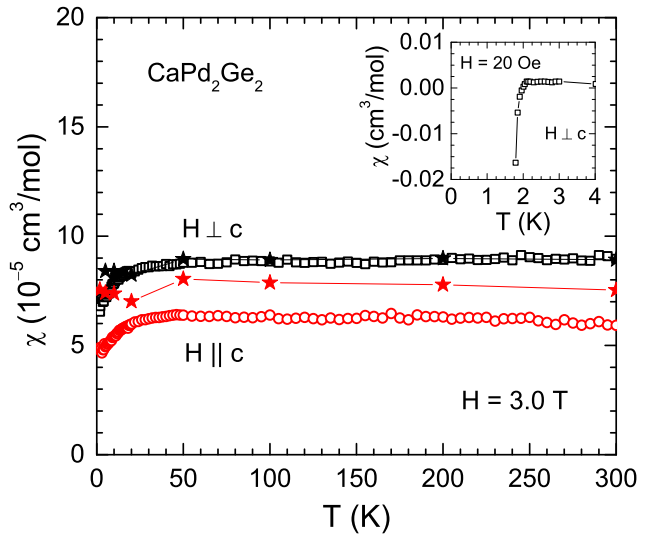}
\caption{ (Color online) Zero-field-cooled magnetic susceptibility $\chi$ of a ${\rm CaPd_2Ge_2}$ single crystal as a function of temperature $T$ over the temperature range 1.8~K~$\leq T \leq300$~K measured in a magnetic field $H = 3.0$~T applied along the $c$ axis ($\chi_c,\ H \parallel {\bf c}$) and in the $ab$~plane ($\chi_{ab},\ H \perp  {\bf c}$). The filled stars represent the intrinsic $\chi$ obtained from fitting $M(H)$ isotherm data by Eq.~(\ref{eq:MH_linear-fit}). The lines joining the stars are guides to the eye.  Inset: Expanded plot of $\chi(T)$ for $T<4$~K measured at the low field $H=20$~Oe showing the onset of superconductivity at $\approx2$~K.}
\label{fig:MT_CaPd2Ge2}
\end{figure}

The zero-field-cooled magnetic susceptibilities $\chi(T) \equiv M(T)/H$ data of a ${\rm CaPd_2Ge_2}$ crystal in the temperature range $1.8~{\rm K}\leq T \leq300$~K for both $H$ applied along the $c$~axis ($\chi_c,\ H \parallel c$) and in the $ab$~plane ($\chi_{ab},\ H \perp c$) are shown in Fig.~\ref{fig:MT_CaPd2Ge2}. The $\chi(T)$ measured at low $H$ (20~Oe) shown in the inset of Fig.~\ref{fig:MT_CaPd2Ge2} reveals the onset of superconductivity at $\approx 2.0$~K, consistent with $T_{\rm c\,onset} = 1.98$~K obtained from the $\rho(T)$ measurements above. The $\chi(T)$ data measured at $H = 3.0$~T exhibit nearly $T$-independent weak paramagnetism for both directions of $H$\@. The downturns in $\chi$ below 25~K arise due to an error in measuring the sample holder contribution which was about 85\% of the total measured moment due to the small mass of the measured crystal.  No such downturns below 25~K are seen in the intrinsic $\chi$ values obtained below from the high-field slopes of $M(H)$ isotherms which are shown by filled stars in Fig.~\ref{fig:MT_CaPd2Ge2}, indicating that the low-$T$ downturns in $\chi(T)$ are extrinsic.  The $\chi$ is weakly anisotropic with $\chi_{ab} > \chi_{c}$ over the entire temperature range $2~{\rm K} \leq T \leq 300$~K, which is the same anisotropy as in the parent $A{\rm Fe_2As_2}$ compounds.\cite{Johnston2010}

\begin{figure}
\includegraphics[width=3.3in]{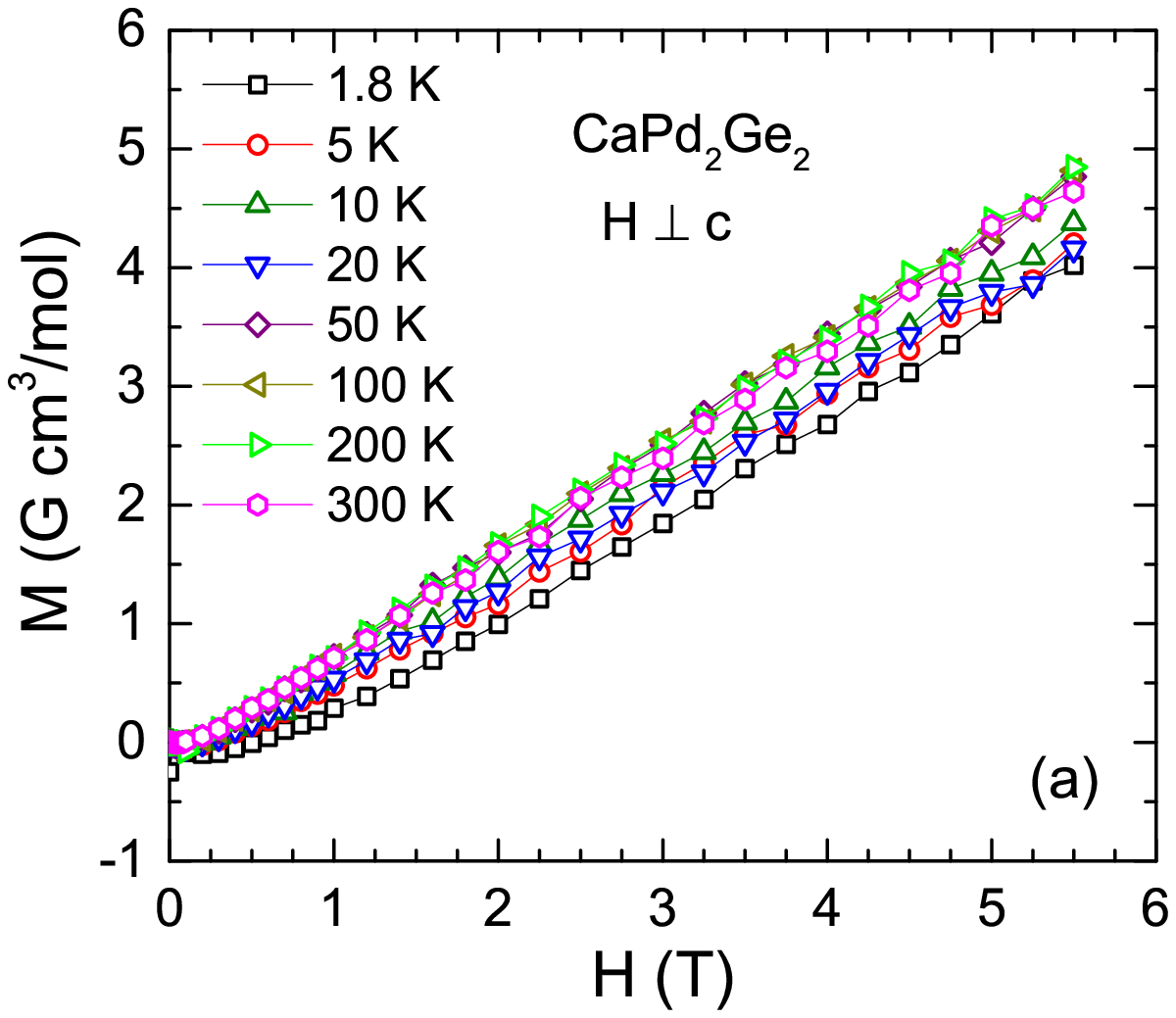}
\includegraphics[width=3.3in]{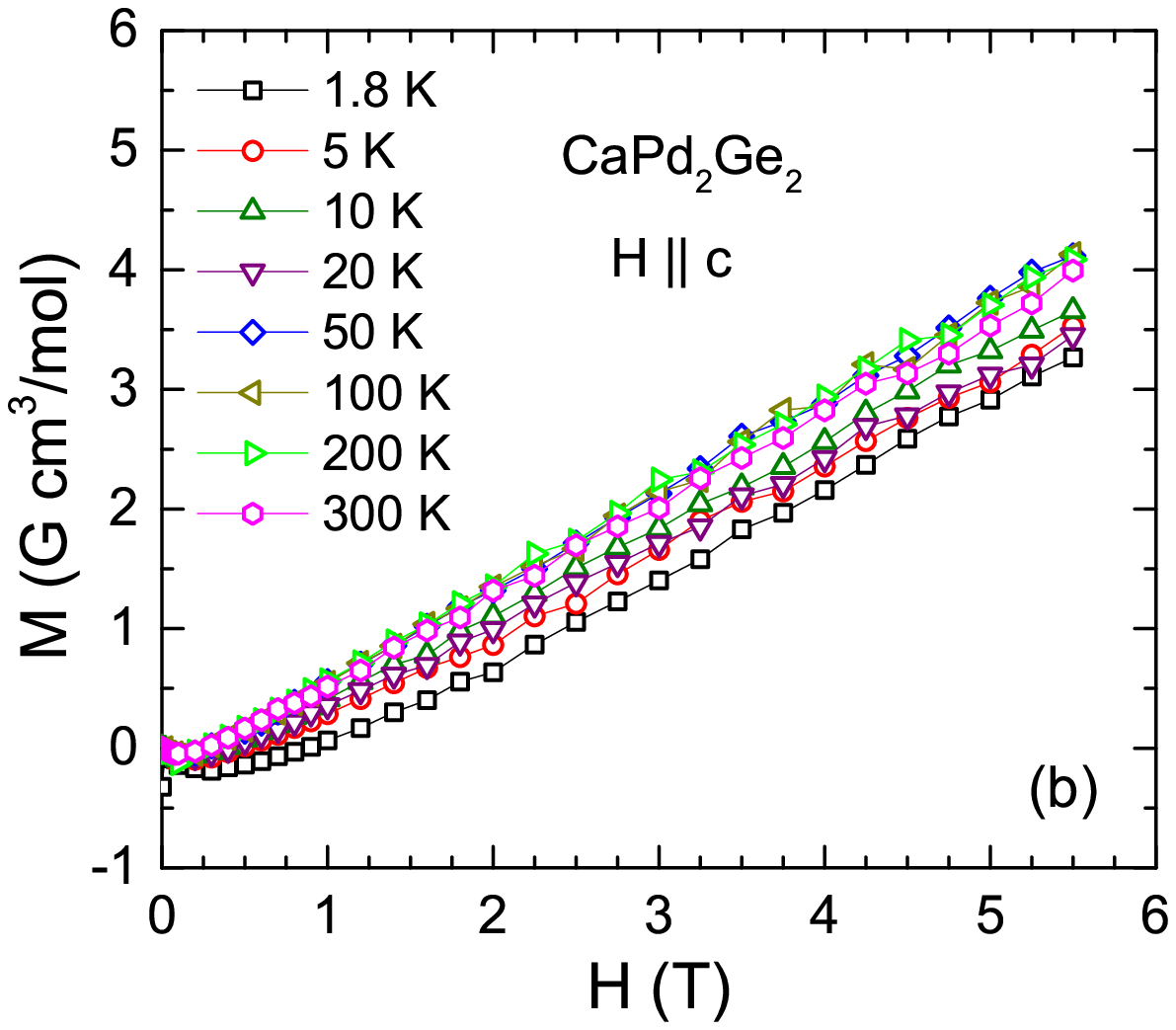}
\caption{(Color online) Isothermal magnetization $M$ of ${\rm CaPd_2Ge_2}$ as a function of applied magnetic field $H$ at different temperatures, as listed, for magnetic fields applied (a) in the $ab$ plane ($H \perp  c$) and, (b) along the $c$ axis ($H \parallel c$).}
\label{fig:MH_CaPd2Ge2}
\end{figure}

The $M(H)$ isotherms of the ${\rm CaPd_2Ge_2}$ crystal for both $H$ applied along the $c$~axis ($M_c,\ H \parallel c$) and in the $ab$~plane ($M_{ab},\ H \perp  c$) measured at eight temperatures between 1.8 and 300~K are shown in Fig.~\ref{fig:MH_CaPd2Ge2}. The $M$ is weakly paramagnetic and anisotropic with $M_{ab}(H) > M_{c}(H)$. The $M(H)$ curves are almost linear in $H$ except at low-$H$\@. In order to obtain the intrinsic $\chi $ we fitted the high-field $M(H)$ data by
\begin{equation}
M(H) = M_{\rm s} + \chi H,
\label{eq:MH_linear-fit}
\end{equation}
where $M_{\rm s}$ is the impurity saturation magnetization. The fitted $\chi$ values obtained from the slope of linear fits of $M(H)$ isotherms for $H \geq 2$~T are shown as filled stars in Fig.~\ref{fig:MT_CaPd2Ge2}. The $M_{\rm s}$ values (not shown) are found to be negative which is unphysical and can be attributed to a small error in the calibration of the sample holder contribution to the total magnetization.

The powder- and temperature-average $\langle\chi\rangle$ of the intrinsic~$\chi$ for $50 \leq T \leq 300$~K in Fig.~\ref{fig:MT_CaPd2Ge2} is $\langle\chi\rangle = [2 \langle\chi_{ab}\rangle + \langle\chi_{c}\rangle]/3 = 8.56\times 10^{-5}$~cm$^3$/mol. The intrinsic $\chi$ contains the four contributions
\begin{equation}
\chi=\chi_{\rm core}+\chi_{\rm VV}+\chi_{\rm L} + \chi_{\rm P}.
\label{eq:chi}
\end{equation}
The first three contributions are orbital terms and the last term is the Pauli spin susceptibility of the conduction carriers.  $\chi_{\rm core}$ is the diamagnetic core susceptibility, $\chi_{\rm VV}$ is the paramagnetic Van Vleck susceptibility and $\chi_{\rm L}$ is the Landau diamagnetic susceptibility of the conduction carriers. Using the constituent atomic diamagnetic susceptibilities,\cite{Mendelsohn1970} we obtain $\chi_{\rm {core}} = -1.79 \times 10^{-4}$~cm$^3$/mol. The $\chi_{\rm P}$ is estimated using the relation \cite{Ashcroft1976} $\chi_{\rm {P}} =  (g^2/2)\mu_{\rm B}^2 {\cal D}_{\rm band}(E_{\rm F})$, where $g$ is the spectroscopic splitting factor ($g$-factor) of the conduction carriers, $\mu_{\rm B}$ is the Bohr magneton, and ${\cal D}_{\rm band}(E_{\rm F})$ is the band-structure density of states at the Fermi energy $E_{\rm F}$.  Using $g=2$ and ${\cal D}_{\rm band}(E_{\rm F}) = 2.47$~states/eV\,f.u. obtained in Eq.~(\ref{Eq:DbandEF}) below from heat capacity measurements, one obtains $\chi_{\rm {P}} = 7.98 \times 10^{-5}$~cm$^3$/mol. The $\chi_{\rm L}$ is estimated from $\chi_{\rm P}$ using the relation \cite{Ashcroft1976, Elliott1998} $\chi_{\rm L} = - (m_{\rm e}/m_{\rm band}^\ast )^2 \chi_{\rm {P}}/3$, which assuming the band structure effective mass $m_{\rm band}^\ast = m_{\rm e}$ gives $\chi_{\rm {L}} = -2.66 \times 10^{-5}$~cm$^3$/mol, where $m_{\rm e}$ is the free-electron mass.  The powder-averaged $\chi_{\rm VV}$ value $\langle\chi_{\rm VV}\rangle$ is then estimated from Eq.~(\ref{eq:chi}) using the above-estimated $\chi_{\rm core}$, $\chi_{\rm P}$, $\chi_{\rm L}$, and $\langle\chi\rangle$ values as $\langle\chi_{\rm VV}\rangle = 2.11\times 10^{-4}$~cm$^3$/mol.

\section{\label{Sec:CaPd2Ge2_HC} Heat Capacity}

\subsection{Overview of the Superconducting- and Normal-State Heat Capacities}

\begin{figure}
\includegraphics[width=3.3in]{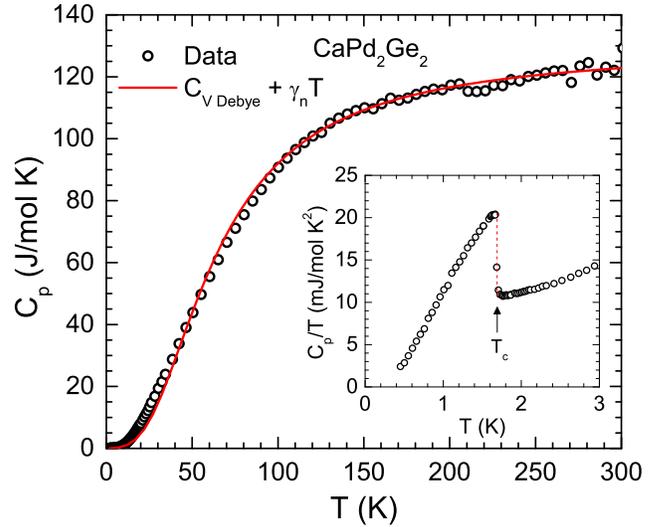}
\caption{(Color online) Heat capacity $C_{\rm p}$ of a ${\rm CaPd_2Ge_2}$ single crystal as a function of temperature $T$ for 1.8~K~$\leq T \leq$~300~K measured in zero magnetic field. The red solid curve is the fitted sum of the contributions from the Debye lattice heat capacity $C_{\rm V\,Debye}(T)$ and predetermined electronic heat capacity $\gamma_{\rm n} T$. Inset: $C_{\rm p}/T$ vs. $T$ for $0.45~{\rm K} \leq T \leq 3.0$~K\@. The dotted red line mark the superconducting transition temperature $T_{\rm c}$.}
\label{fig:CaPd2Ge2_HC}
\end{figure}

\begin{figure}
\includegraphics[width=3.3in]{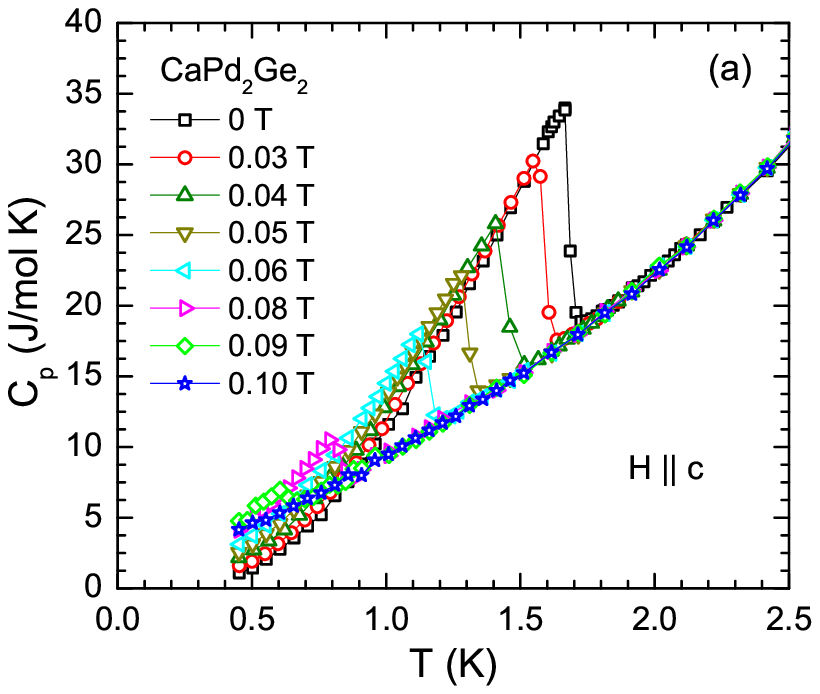}
\includegraphics[width=3.3in]{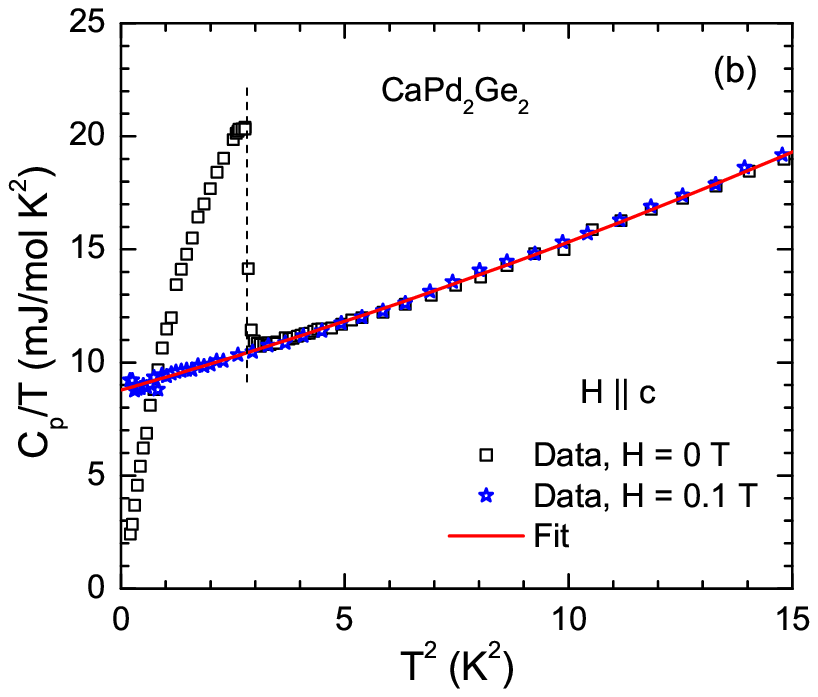}
\caption{(Color online) (a) Heat capacity $C_{\rm p}$ as a function of temperature $T$ of a ${\rm CaPd_2Ge_2}$ single crystal for 0.45~K~$\leq T \leq$~2.5~K measured in different magnetic fields $H$ applied  along the $c$ axis as indicated. (b) $C_{\rm p}/T$ versus $T^2$ for 0.45~K~$\leq T \leq$~3.9~K with $H = 0$ and 0.1~T\@. The red curve is a fit of the $H = 0.1$~T data for $0.45~{\rm K} \leq T \leq 4.0$~K together with the $H = 0$ data for $1.8~{\rm K}\leq T \leq 4.0$~K by Eq.~(\ref{Eq:CpTFit}).}
\label{fig:CaPd2Ge2_HC_field}
\end{figure}

The $C_{\rm p}(T)$ data of ${\rm CaPd_2Ge_2}$ measured at $H = 0$ in 0.45~K~$\leq T \leq$~300~K are shown in Fig.~\ref{fig:CaPd2Ge2_HC}. A sharp jump in $C_{\rm p}$ due to the occurrence of superconductivity at $T_{\rm c} = 1.69(3)$~K is evident from the low-$T$ $C_{\rm p}(T)$ data shown in the inset of Fig.~\ref{fig:CaPd2Ge2_HC}. The jump in $C_{\rm p}$ and $T_{\rm c}$ are found to decrease with the application of magnetic field as can be seen from the $C_{\rm p}(T)$ data measured at various $H$ shown in Fig.~\ref{fig:CaPd2Ge2_HC_field}(a). The $T_{\rm c}$ is found to suppress to below 0.45~K by $H = 0.10$~T\@.

Plots of $C_{\rm p}/T$ vs. $T^2$ for $H = 0$ and 0.10~T are shown in Fig.~\ref{fig:CaPd2Ge2_HC_field}(b). In order to estimate the normal-state Sommerfeld electronic heat capacity coefficient $\gamma_{\rm n}$ we fitted the (normal-state) $C_{\rm p}(T)$ data at $H=0.10$~T and the normal-state part of the $C_{\rm p}(T)$ data at $H=0$  simultaneously by
\be
\frac{C_{\rm p}(T)}{T} = \gamma_{\rm n} + \beta T^2 + \delta T^4,
\label{Eq:CpTFit}
\ee
where $\beta T^3$ is the Debye $T^3$-law low-$T$ lattice heat capacity and $\delta T^5$ is an additional lattice heat capacity contribution.  The fit of $C_{\rm p}(T)$ data at $H=0.10$~T in $0.45~{\rm K} \leq T \leq 4.0$~K and at $H=0$ in $1.8~{\rm K}\leq T \leq 4.0$~K is shown by the red curve in Fig.~\ref{fig:CaPd2Ge2_HC_field}(b). We thus obtained $\gamma_{\rm n} = 8.78(4)$~mJ/mol\,K$^2$, $\beta= 0.56(2)$~mJ/mol\,K$^4$ and $\delta = 9.4(8)~\mu$J/mol\,K$^6$. The Debye temperature $\Theta_{\rm D}= 259(3)$~K was estimated from $\beta$ using the relation \cite{Kittel2005} $\Theta_{\rm D} = (12 \pi^{4} R p/5 \beta )^{1/3}$, where $R$ is the molar gas constant and $p=5$ is the number of atoms per formula unit (f.u.).

The high-temperature $C_{\rm p}(T)$ data in Fig.~\ref{fig:CaPd2Ge2_HC} yield $C_{\rm p}(T = 300~{\rm K}) \approx 123$~J/mol\,K, which   is close to the expected classical Dulong-Petit value $C_{\rm V} = 3pR = 15R$ = 124.7~J/mol\,K at constant volume.\cite{Kittel2005, Gopal1966} The normal-state $C_{\rm p}(T)$ data at $H=0$ over the full temperature range 2.0~K~$\leq T \leq$~300~K were fitted by the sum of the electronic term $\gamma_{\rm n} T$ determined above and the Debye model lattice heat capacity $C_{\rm V\,Debye}(T)$,\cite{Goetsch2012, Kittel2005, Gopal1966} yielding the fitted Debye temperature $\Theta_{\rm D}= 259(2)$~K\@.  The fit is shown as the solid red curve in Fig.~\ref{fig:CaPd2Ge2_HC}. This value of $\Theta_{\rm D}$ agrees within the statistical error bars with the value $\Theta_{\rm D}= 259(3)$~K obtained above from the coefficient $\beta$.  This good agreement is very unusual because $\Theta_{\rm D}$ is temperature-dependent due to the difference between the actual phonon density of states versus frequency of a material and the quadratic behavior assumed in the Debye theory for all materials.  On the other hand, this value of $\Theta_{\rm D}$ is much larger than the $\Theta_{\rm{R}} = 171(3)$~K obtained above from the fit of the $\rho(T)$ data by the Bloch-Gr\"uneisen theory.  Often these differently-derived values of the Debye temperature are in fairly good agreement.  It is not clear why they are so different for ${\rm CaPd_2Ge_2}$ since the respective experimental data from both types of measurements are well fitted by the respective theory.

The density of states at the Fermi energy ${\cal D}_C(E_{\rm F})$ determined from heat capacity measurements was estimated from the value of $\gamma_{\rm n}$ according to \cite{Kittel2005}
\begin{equation}
\gamma_{\rm n} = \frac{\pi^2 k_{\rm B}^2}{3}\, {\cal D}_C(E_{\rm F}).
\label{eq:DOS}
\end{equation}
Using the above value $\gamma_{\rm n} = 8.78(4)$~mJ/mol\,K$^2$ gives
\be
{\cal D}_C(E_{\rm F}) = 3.72(2)~{\rm states/eV\,f.u.}
\ee
for both spin directions. This value of ${\cal D}_C(E_{\rm F})$ includes the influence of the electron-phonon interaction. The bare band-structure density of states ${\cal D}_{\rm band}(E_{\rm F})$ and $m^\ast_{\rm band}$ are related to  ${\cal D}_C(E_{\rm F})$ according to\cite{Grimvall1976}
\bse
\bea
{\cal D}_C(E_{\rm F}) &=& {\cal D}_{\rm band}(E_{\rm F})(1 + \lambda_{\rm e-ph}),
\label{eq:DOSband}\\*
m^\ast &=& m^\ast_{\rm band}(1 + \lambda_{\rm e-ph}),\label{Eq:m*}
\eea
\ese
where $\lambda_{\rm {e-ph}}$ is the electron-phonon coupling constant and $m^\ast$ is the effective mass of the quasiparticles (charge carriers).  An estimate of $\lambda_{\rm {e-ph}}$ is made from $\Theta_{\rm D}$ and $T_{\rm c}$ using the relation from McMillan's theory\cite{McMillan1968}
\begin{equation}
\lambda_{\rm {e-ph}}= \frac {1.04+\mu^{\ast} \ln(\Theta_{\rm D}/1.45\,T_{\rm c})} {(1-0.62\mu^{\ast})\ln(\Theta_{\rm D}/1.45\,T_{\rm c}) - 1.04}.
\label{eq:lambda}
\end{equation}
where $\mu^{\ast}$ is the repulsive screened Coulomb parameter. Taking $\mu^{\ast}$ to be the representative value $\mu^{\ast} = 0.13$ and using the values $T_{\rm c} = 1.69$~K and $\Theta_{\rm D}= 259$~K, Eq.~(\ref{eq:lambda}) gives $\lambda_{\rm {e-ph}} = 0.51$. This value of $\lambda_{\rm {e-ph}}$ is rather small and indicates weak-coupling superconductivity in ${\rm CaPd_2Ge_2}$. The value
\be
{\cal D}_{\rm band}(E_{\rm F})= 2.47~{\rm \frac{states}{eV\,f.u.}}
\label{Eq:DbandEF}
\ee
for both spin directions is obtained from the values of ${\cal D}_C(E_{\rm F})$ and $\lambda_{\rm {e-ph}}$ using Eq.~(\ref{eq:DOSband}).

Next we estimate the mean free path $\ell$ from the values of ${\cal D}_C(E_{\rm F})$ and the residual resistivity $\rho_0$. The Fermi velocity $v_{\rm F}$ is related to ${\cal D}_C(E_{\rm F})$ by \cite{Kittel2005}
\begin{equation}
v_{\rm F} = \frac {\pi^2 \hbar^3}{{m^{\ast}}^2 V_{\rm f.u.}} {\cal D}_C(E_{\rm F}),
\label{eq:vF}
\end{equation}
where $V_{\rm f.u.} = V_{\rm cell}/2$ is the volume per formula unit and $\hbar$ is Planck's constant divided by $2\pi$. This gives $v_{\rm F}  = 1.53 \times 10^8~{\rm cm/s}$ where we assumed that the band mass is $m^\ast_{\rm band} = m_{\rm e}$ and hence $m^\ast =1.53 m_{\rm e}$ from Eq.~(\ref{Eq:m*}). The mean free path is given by $\ell = v_{\rm F}\tau$, where the mean free scattering time $\tau$ is related to $\rho_0$ by $\tau= m^\ast /(n e^2 \rho_0)$, \cite{Kittel2005} $n$ being the conduction carrier density given by \cite{Kittel2005}
\be
n = \frac{1}{3\pi^2}\left(\frac{m^\ast v_{\rm F}}{\hbar}\right)^3.
\ee
Thus
\be
\ell = 3\pi^2 \left(\frac{\hbar}{e^2\rho_0}\right)\left(\frac{ \hbar}{m^\ast v_{\rm F}}\right)^2,
\label{eq:lvF}
\ee
where $\hbar/e^2 = 4108~\Omega$. Using the above estimated value of $v_{\rm F}$ for $\rho_0 = 12~\mu\Omega$\,cm we obtain $\ell = 2.55~{\rm nm}$, where again we used $m^\ast = 1.51m_{\rm e}$.

The superconducting London penetration depth in the clean limit at $T=0$, $\lambda_{\rm L}(0)$, is a normal-state property since it can be estimated from the Fermi velocity using the relation \cite{Tinkham1996}
\be
\lambda_{\rm L}(0) = \frac{c}{\omega_{\rm p}},
\label{Eq:lambda0}
\ee
where $c$ is the speed of light in vacuum and the plasma angular frequency $\omega_{\rm p}$ of the conduction carriers is given by
\be
\omega_{\rm p}^2 = \frac{4\pi n e^2}{m^*} = \frac{4(m^\ast e)^2(v_{\rm F}/\hbar)^3}{3\pi}.
\label{Eq:omegapDef}
\ee
We thus obtain $\omega_{\rm p} = 2.37\times10^{16}~{\rm rad/s}$ from Eq.~(\ref{Eq:omegapDef}) and $\lambda_{\rm L}(0) = 12.6~{\rm nm}$ from Eq.~(\ref{Eq:lambda0}) using $m^\ast=1.51m_{\rm e}$ and $v_{\rm F}  = 1.53 \times 10^8~{\rm cm/s}$.

The above normal-state parameters are summarized in Table~\ref{tab:SCParams}.

\begin{table}
\caption{\label{tab:SCParams} Measured and derived superconducting and relevant normal state parameters for ${\rm CaPd_2Ge_2}$.  $T_{\rm c}$: bulk superconducting transition temperature; $\gamma_{\rm n}$: observed Sommerfeld coefficient of the linear term in the low-$T$ normal-state heat capacity; ${\cal D}(E_{\rm F})$: density of states at the Fermi energy for both spin directions determined from $\gamma_{\rm n}$; $\Theta_{\rm D}$: Debye temperature from fitting heat capacity data; $\Theta_{\rm R}$: Debye temperature from fitting electrical resistivity data; $\lambda_{\rm e-ph}$: electron-phonon coupling constant; $\Delta C_{\rm e}$: heat capacity jump at $T_{\rm c}$; $\alpha = \Delta(0)/k_{\rm B}T_{\rm c}$; $\Delta$: superconducting order parameter determined from London penetration depth measurements; $\alpha_{\rm M}$: Maki parameter; $H_{\rm c}$, $H_{\rm P}$, $H_{\rm c1}$, $H_{\rm c2}^{\rm Orb}$, $H_{\rm c2}$: thermodynamic, Paul limiting upper critical, lower critical, orbital upper critical, and fitted upper critical magnetic fields, respectively; $\kappa_{\rm GL}$: Ginzburg-Landau parameter; $\xi$: Ginzburg-Landau coherence length; $\xi_0$: BCS superconducting coherence length; $\ell$: electronic mean-free path at low~$T$; $\omega_{\rm p}$: angular plasma frequency; $\lambda_{\rm L}$: London penetration depth; $\lambda_{\rm eff}$: magnetic penetration depth. The value of $\lambda^{\rm obs}(0)$ is determined from the London penetration depth measurements.}
\begin{ruledtabular}
\begin{tabular}{lcc}
CaPd$_2$Ge$_2$ property & value\\
\hline
$T_{\rm c}$ (K)                                         		&  1.69(3)   \\
$\gamma_{\rm n}$ (mJ/mol\,K$^{2}$)                      		& 8.78(4) \\
${\cal D}_C(E_{\rm F})$ (states/eV\,f.u.)					& 3.72(2)		\\
$\Theta_{\rm D}$ (K) from low-$T$ $C_{\rm p}$ data      		& 259(3) \\
$\Theta_{\rm D}$ (K) from all-$T$ $C_{\rm p}$ data      		& 259(2) \\
$\Theta_{\rm R}$ (K) from $\rho(T)$ data      				& 171(3) \\
$\lambda_{\rm e-ph}$                                    		& 0.51   \\
$\Delta C_{\rm e}$ (mJ/mol\,K)                          		& 17.9(3) \\
$\Delta C_{\rm e}/\gamma_{\rm n} T_{\rm c}$             		& 1.21(3) \\
$\alpha$ (from $\Delta C_{\rm e}/\gamma_{\rm n} T_{\rm c}$) 	& 1.62(3)\\
$\Delta(0)/k_{\rm B}$~(K) (observed)				    		& 2.62(5)\\
$\alpha_{\rm M}$                                        		&  0.065 \\
$H_{\rm c}(T=0)$ (mT)                                   		&  15.0 \\
$H_{\rm P}$ (T) 									    	&  2.89\\
$H_{\rm c1}(T=0)$ (mT)                                  		&  3.1   \\
$H_{\rm c2}^{\rm Orb}(T=0)$~(dirty limit) (T)           		& 0.13(2) \\
$H_{\rm c2}(T=0)$ (mT)                                  		& 134 \\
$\kappa_{\rm GL}$                                       		&  6.30\\
$\xi(T=0)$ (nm)                                         		&  49.7  \\
$\xi_0$ (nm)                                            		&  1181     \\
$\ell~(m^\ast=1.51m_{\rm e})$ (nm)                          	&  2.55    \\
$\omega_{\rm p}~(m^\ast=1.51m_{\rm e})~(10^{16}~{\rm rad/s})$ 	& 2.37 \\
$\lambda_{\rm L}^{\rm calc}(0)$~(clean limit) (nm)      		&  12.6   \\
$\lambda_{\rm eff}^{\rm calc}(0)$~(dirty limit) (nm)    		&  272--313 \\
$\lambda^{\rm obs}(0)$ (nm)              		       		&  186(16)    \\
\end{tabular}
\end{ruledtabular}
\end{table}

\subsection{Superconducting-State Properties}

\begin{figure}
\includegraphics[width=3.3in]{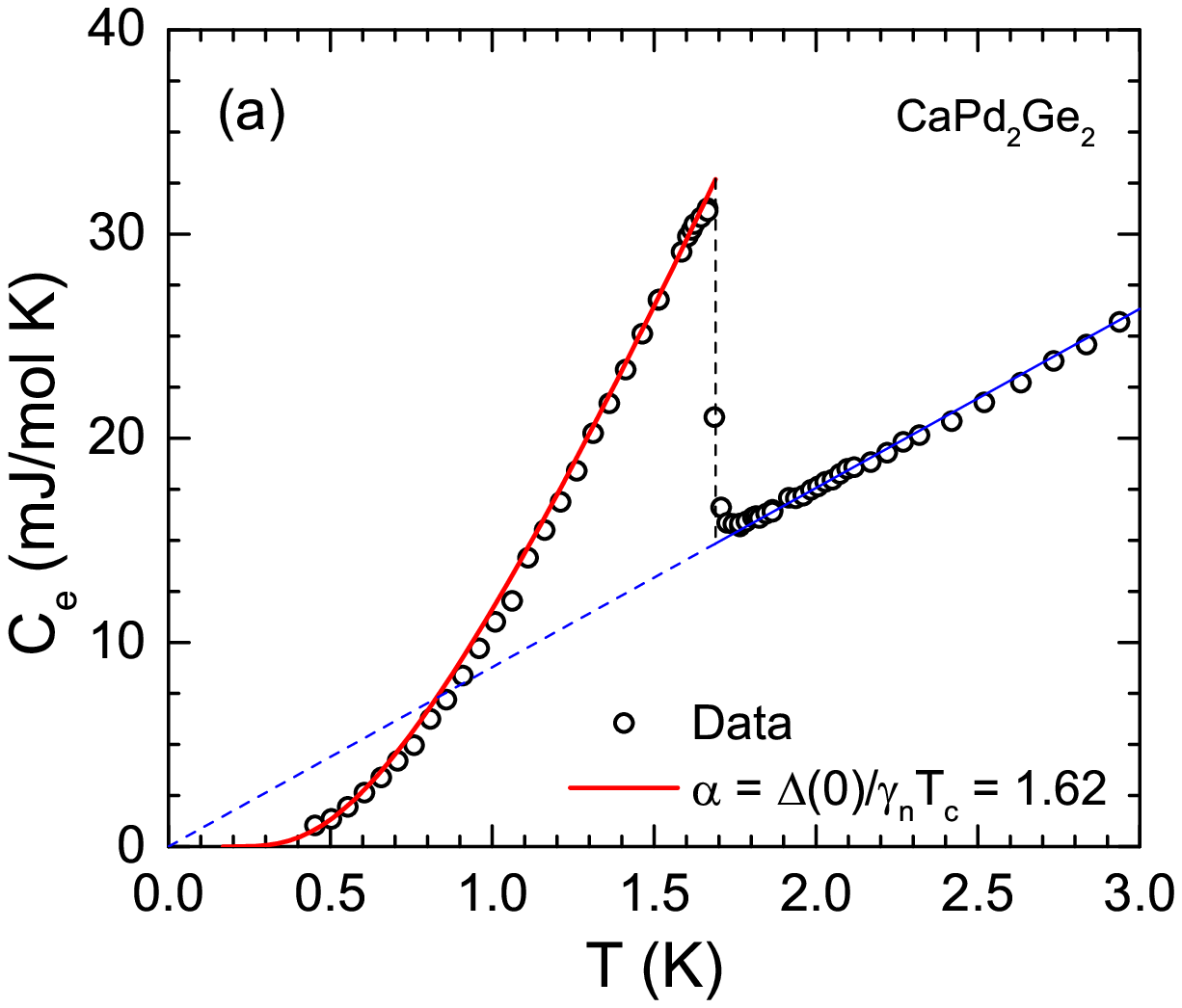}
\includegraphics[width=3.3in]{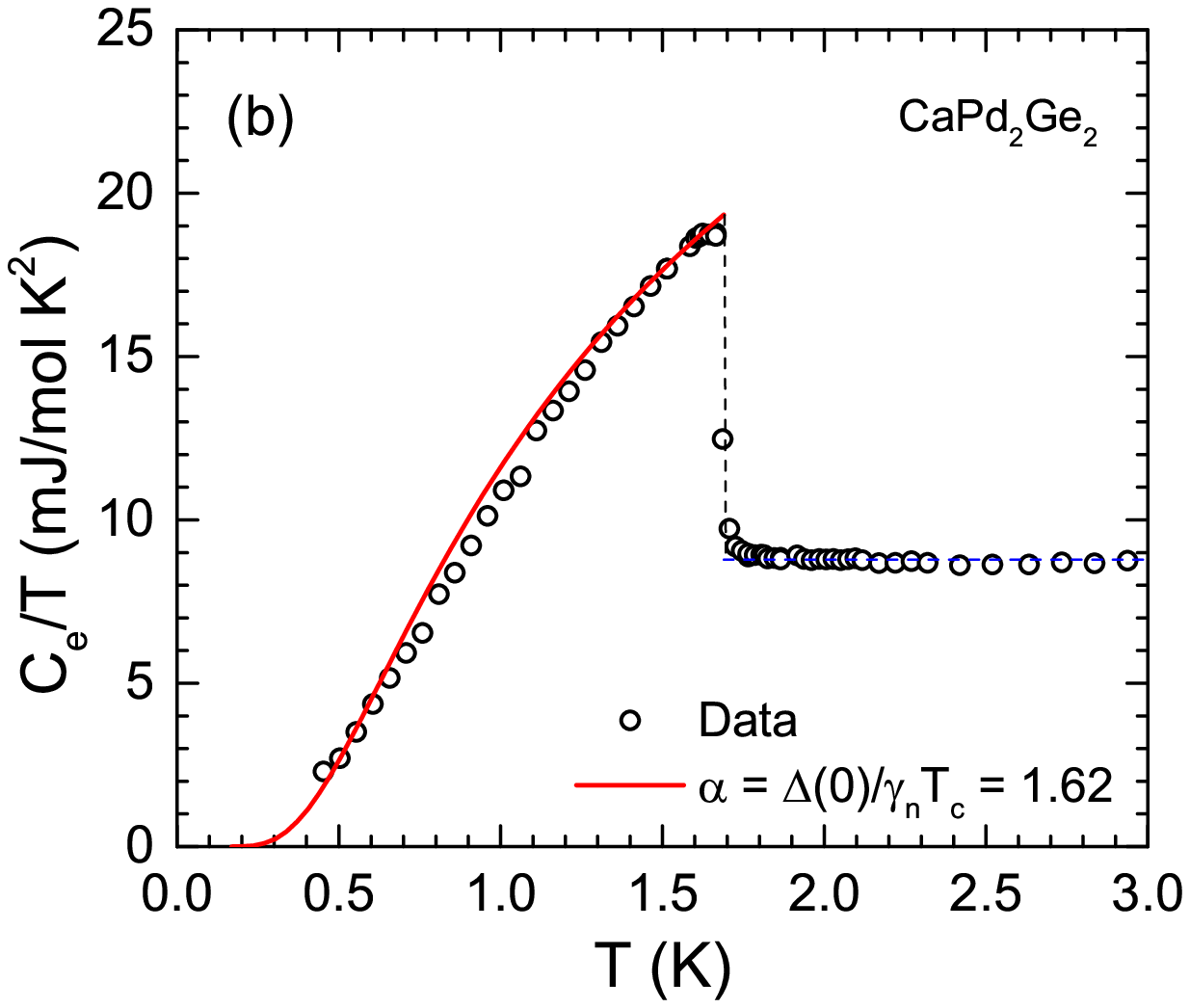}
\caption{(Color online) (a) Electronic contribution $C_{\rm e}$ to the heat capacity as a function of temperature $T$ of ${\rm CaPd_2Ge_2}$.  The normal-state behavior is shown by the straight full blue line and the extrapolation to $T=0$ by the dashed blue line.  (b) $C_{\rm e}/T$ versus $T$\@. The solid red curve in each figure is the theoretical prediction of the $\alpha$-model for $\alpha = 1.62$.  The BCS value is $\alpha_{\rm BCS} \approx 1.764$.}
\label{fig:CaPd2Ge2_HC_el}
\end{figure}

The electronic contribution to the heat capacity $C_{\rm e}(T)$ for ${\rm CaPd_2Ge_2}$ is shown in Fig.~\ref{fig:CaPd2Ge2_HC_el}. The $C_{\rm e}(T)$ was obtained by subtracting the above-determined phonon contribution from the measured $C_{\rm p}(T)$ data, i.e., $C_{\rm e}(T) = C_{\rm p}(T) - (\beta T^3 + \delta T^5$). A sharp jump $\Delta C_{\rm e}$ in $C_{\rm e}$ at $T_{\rm c}$ reveals bulk superconductivity in ${\rm CaPd_2Ge_2}$. From Fig.~\ref{fig:CaPd2Ge2_HC_el}, $\Delta C_{\rm e} = 17.9(3)$~mJ/mol\,K or $\Delta C_{\rm e}/T_{\rm c} = 10.6(3)~{\rm mJ/mol\,K^2}$ is obtained corresponding to the entropy-conserving construction shown by the vertical dotted line at $T_{\rm c}$ in Fig.~\ref{fig:CaPd2Ge2_HC_el}(a). The normalized value $\Delta C_{\rm e}/ \gamma_{\rm n} T_{\rm c} = 1.21(3)$ is obtained using the above estimated $\gamma_{\rm n} = 8.78(4)$~mJ/mol\,K$^2$. This intrinsic value of $\Delta C_{\rm e}/ \gamma_{\rm n} T_{\rm c}$ is significantly smaller than the BCS value $\Delta C_{\rm e}/ \gamma_{\rm n} T_{\rm c} =1.426$.\cite{Tinkham1996}  We attribute this reduction to the presence of an anisotropic superconducting energy gap in a single $s$-wave gap model\cite{Johnston2013} and use the single-band $\alpha$-model of the BCS theory superconductivity \cite{Bardeen1957, Padamsee1973, Johnston2013} for further analysis of the superconducting state data.

In the $\alpha$-model, the BCS parameter $\alpha_{\rm BCS} \equiv \Delta(0)/k_{\rm B}T_{\rm c} = 1.764$ is used to calculate the order parameter $\Delta(T)$ and the London penetration depth, but is replaced by a variable $\alpha$ and hence a variable $\Delta(0)/k_{\rm B}T_{\rm c}$ in fitting other properties such as the jump in the electronic heat capacity at $T_{\rm c}$.\cite{Padamsee1973, Johnston2013}  Using the relation \cite{Johnston2013}
\be
\frac{\Delta C_{\rm e}(T_{\rm c})}{\gamma_{\rm n}T_{\rm c}} = 1.426\left(\frac{\alpha}{\alpha_{\rm BCS}}\right)^2,
\ee
for ${\rm CaPd_2Ge_2}$ we obtain $\alpha = 1.62(3)$ which is significantly smaller than the BCS value of~1.764. The solid red curves in Figs.~\ref{fig:CaPd2Ge2_HC_el}(a) and \ref{fig:CaPd2Ge2_HC_el}(b) are the theoretical predictions of the $\alpha$-model for $C_{\rm e}(T)$ using $\alpha = 1.62$. The predicted superconducting state $C_{\rm es}(T)$ and $C_{\rm es}(T)/T$ behaviors are seen to be in good agreement with the $T$-dependent experimental data. Additional information on fitting $C_{\rm es}(T)$ data within the framework of the $\alpha$-model is given in Refs.~\onlinecite{Anand2013a} and \onlinecite{Johnston2013}.

\begin{figure}
\includegraphics[width=3.3in]{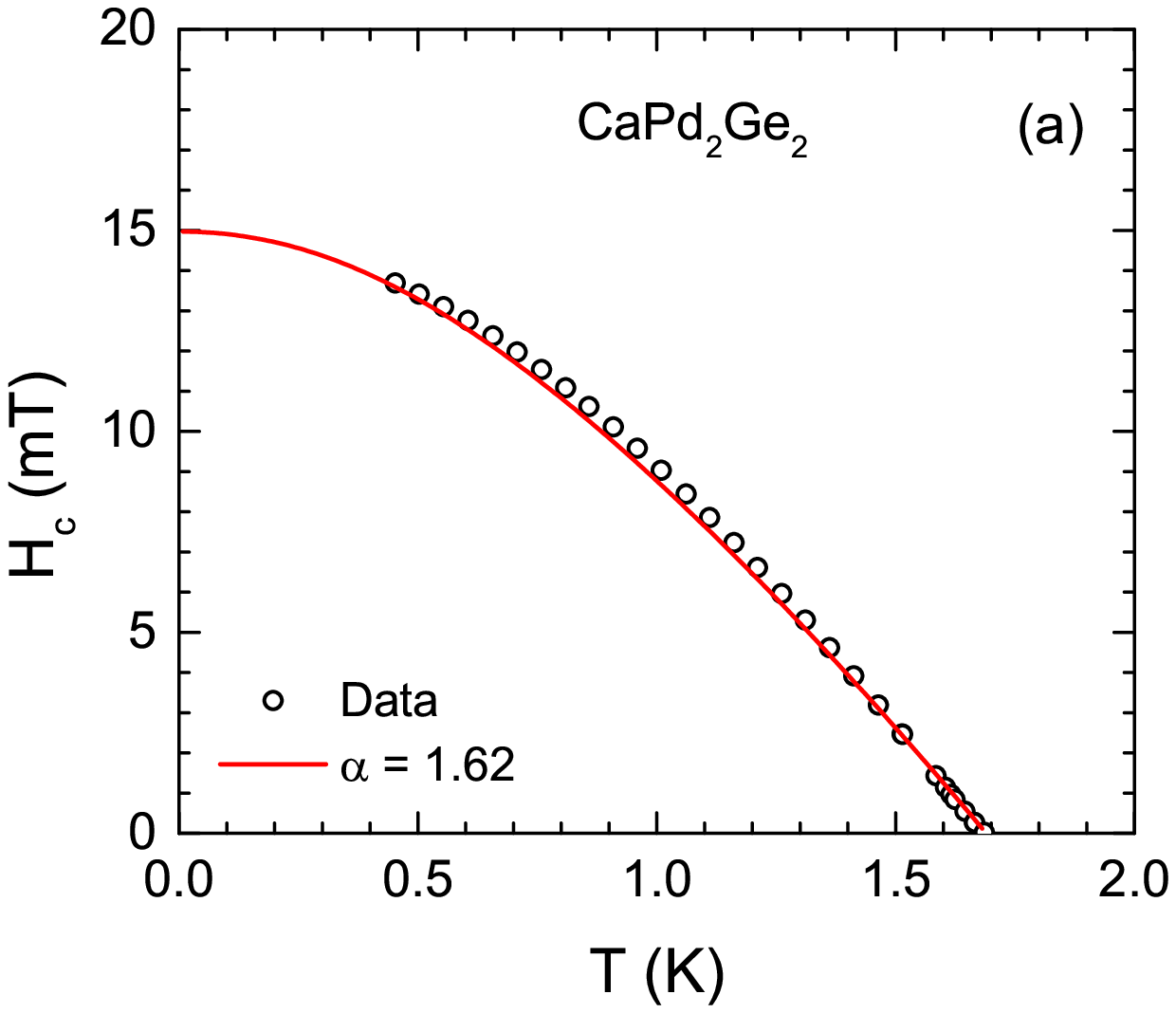}
\includegraphics[width=3.3in]{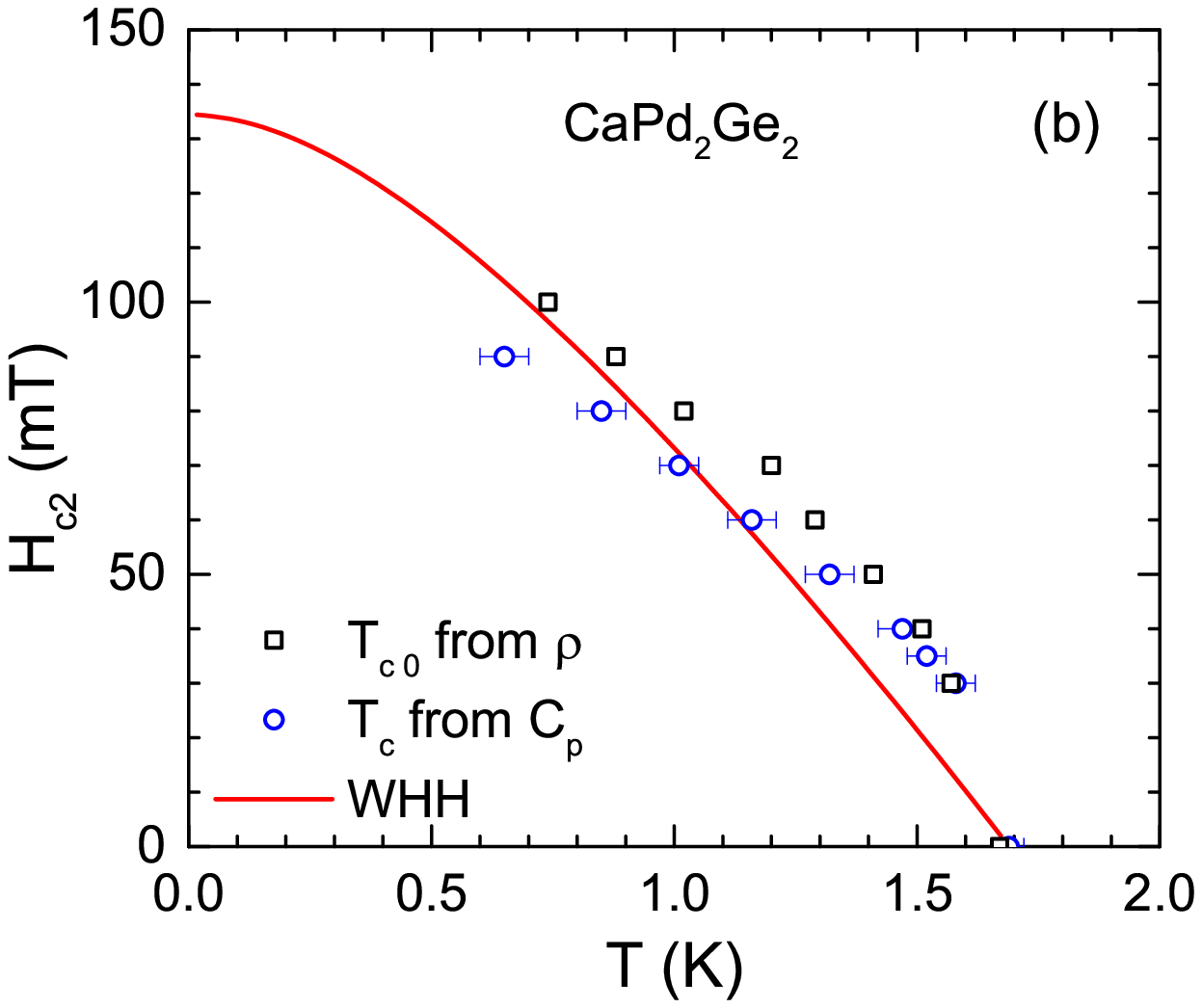}
\caption{(Color online) (a) Thermodynamic critical field $H_{\rm c}$ as a function of temperature~$T$ obtained for ${\rm CaPd_2Ge_2}$ from the experimentally derived electronic heat capacity $C_{\rm e}(T)$ data using Eq.~(\ref{Eq:HcFromCp}). The solid red curve is the theoretical prediction of the $\alpha$-model\cite{Padamsee1973, Johnston2013} for $T_{\rm c} = 1.69~K$, $\alpha=1.62$ and $\gamma_{\rm n} = 8.78~{\rm mJ/mol\,K^2}$. (b) Upper critical magnetic field $H_{\rm c2}(T)$ of ${\rm CaPd_2Ge_2}$ determined from the electrical resistivity $\rho(T, H)$ (black open squares)  and heat capacity $C_{\rm p}(T, H)$ (blue open circles) data in Figs.~\ref{fig:CaPd2Ge2_Rho}(b) and \ref{fig:CaPd2Ge2_HC_field}(a), respectively. The designation $T_{\rm c\,0}$ is the temperature at which zero resistivity is reached on cooling. The red curve is the prediction for $H_{\rm c2}(T)$  by the WHH theory\cite{WHH1966} in Eqs.~(\ref{Eqs:WHHEqns}) for~$\alpha_{\rm M}=0.115$ and~$\lambda_{\rm so}=0$.}
\label{fig:CaPd2Ge2_critical_H}
\end{figure}

Within the $\alpha$-model, the thermodynamic critical field $H_{\rm c}$ at $T=0$ is given by \cite{Johnston2013}
\be
\frac{H_{\rm c}(0)}{\left(\gamma_{\rm nV}T_{\rm c}^2\right)^{1/2}} = \sqrt{\frac{6}{\pi}}\ \alpha \approx 1.382\,\alpha,
\label{Eq:HcFromGammaTc}
\ee
where the Sommerfeld coefficient per unit volume $\gamma_{\rm nV}$ is in units of ${\rm erg/(cm^3\,K^2)}$, and for our compound ${\rm CaPd_2Ge_2}$ we obtain $\gamma_{\rm nV} = 1566~{\rm erg/cm^3\,K^2}$ from the above experimental value of $\gamma_{\rm n}$ and the molar volume.  Then using $\alpha = 1.62$ and $T_{\rm c} = 1.69$~K, Eq.~(\ref{Eq:HcFromGammaTc}) gives $H_{\rm c}(0)= 15.0$~mT\@.

The experimental thermodynamic $H_{\rm c}(T)$ is determined from the zero-field $C_{\rm e}(T)$ data according to \cite{Tinkham1996, DeGennes1966}
\be
H_{\rm c}^{2}(T) = 8\pi\int_{T}^{T_{\rm c}}[S_{\rm en}(T^\prime)-S_{\rm es}(T^\prime)] dT^\prime.
\label{Eq:HcFromCp}
\ee
where the electronic contribution to the entropy is $S_{\rm e}(T^{\prime}) = \int_0^{T^\prime}[C_{\rm e}(T^{\prime\prime})/T^{\prime\prime})]dT^{\prime\prime}$, and $S_{\rm en}$ and $S_{\rm es}$ are the electronic entropies of the normal and superconducting states, respectively. The experimental $H_{\rm c}(T)$ obtained from the zero-field $C_{\rm e}(T)$ according to Eq.~(\ref{Eq:HcFromCp}) is shown in Fig.~\ref{fig:CaPd2Ge2_critical_H}(a). The $\alpha$-model prediction for $H_{\rm c}(T)$ is shown by the solid red curve in Fig.~\ref{fig:CaPd2Ge2_critical_H}(a). The $H_{\rm c}(T)$ data and the theoretical prediction of the $\alpha$-model are seen to be in very good agreement.

The $T$ dependences of the upper critical field $H_{\rm c2}$ of ${\rm CaPd_2Ge_2}$ obtained from the $\rho(T)$ data in Fig.~\ref{fig:CaPd2Ge2_Rho}(b) at which zero resistance is reached on cooling and the $C_{\rm p}(T)$ data in Fig.~\ref{fig:CaPd2Ge2_HC_field}(a) measured with $H \parallel c$~axis are shown in Fig.~\ref{fig:CaPd2Ge2_critical_H}(b). It is seen that the $H_{\rm c2}(T\to0)$ values are much larger than the thermodynamic critical field $H_{\rm c}(0) = 15.0$~mT determined above which shows that ${\rm CaPd_2Ge_2}$ is a type-II superconductor.

The orbital critical field at $T=0$, $H_{\rm c2}^{\rm Orb}(0)$, is related to the initial slope $dH_{\rm c2}(T)/dT|_{T=T_{\rm c}}$ of $H_{\rm c2}(T)$ according to\cite{Hefland1966,WHH1966}
\be
H_{\rm c2}^{\rm Orb}(0) = - A\,T_{\rm c} \frac{dH_{\rm c2}(T)}{dT}\bigg|_{T=T_{\rm c}},
\label{Eq:Hc2Orb}
\ee
where $A = 0.73$ and~0.69 in the clean and dirty limits, respectively. Within the $\alpha$-model of the BCS theory of superconductivity, the Pauli-limiting upper critical field at $T=0$, $H_{\rm P}(0)$, is given by \cite{Johnston2013}
\be
H_{\rm P}(0)[{\rm T}] = 1.86\,T_{\rm c}[{\rm K}]\left(\frac{\alpha}{\alpha_{\rm BCS}}\right),
\label{Hp(0)}
\ee
which for $T_{\rm c} = 1.69$~K, $\alpha = 1.62$ obtained above and $\alpha_{\rm BCS} = 1.764$ gives $H_{\rm P}(0) = 2.89$~T\@. The measured $H_{\rm c2}(0)$ is much smaller than $H_{\rm P}(0)$, indicating that the effect of Pauli limiting on $H_{\rm c2}$ is weak.

We further analyze the $H_{\rm c2}(T)$ data using the dirty-limit  WHH (Werthamer, Helfand and Hohenberg) theory. \cite{WHH1966} We confirm below that ${\rm CaPd_2Ge_2}$ is in the dirty limit.  The WHH theory includes the influence of both Pauli limiting and spin-orbit scattering of quasiparticles on $H_{\rm c2}$ according to \cite{WHH1966}
\bse
\label{Eqs:WHHEqns}
\begin{eqnarray}
\ln\frac{1}{t} & = &\sum_{\nu=-\infty}^\infty \Bigg\{ \frac{1}{|2\nu+1|} - \bigg[ |2\nu+1| + \frac{\bar{h}}{t}  \label{eq:WHH} \\
 && \hspace{0.6in}+\ \frac{(\alpha_{\rm M}\bar{h}/t)^2}{|2\nu+1|+(\bar{h} +\lambda_{\rm so})/t}\bigg] ^{-1}  \Bigg\},
\nonumber
\end{eqnarray}
where the dimensionless $t = T/T_{\rm c}$, $\lambda_{\rm so}$ is the spin-orbit scattering parameter and $\bar{h}$ is defined as
\be
\bar{h} = - \left(\frac{4}{\pi^2}\right) \frac{H_{\rm c2}(T)/T_{\rm c}}{dH_{\rm c2}(T)/dT|_{T=T_{\rm c}}} .
\ee
\ese
The Maki parameter $\alpha_{\rm M}$ is a measure of the relative strengths of the orbital and Pauli-limiting values of $H_{\rm c2}$ and is given by \cite{Maki1966}
\be
\alpha_{\rm M} = \sqrt{2}\ \frac{H_{\rm c2}^{\rm Orb}(0)}{H_{\rm P}(0)}.
\label{Eq:alphaMDef}
\ee
Using Eq.~(\ref{Eq:Hc2Orb}) and $H_{\rm P}(0) = 2.89$~T obtained above,  Eq.~(\ref{Eq:alphaMDef}) yields the Maki parameter as
\be
\alpha_{\rm M} = -0.823\,A\ \frac{dH_{\rm c2}(T)}{dT}\Big|_{T=T_{\rm c}}.
\ee
Taking $A=0.69$ for the dirty limit appropriate to the WHH theory gives
\be
\alpha_{\rm M} = -0.57 \frac{dH_{\rm c2}(T)}{dT}\Big|_{T=T_{\rm c}}.
\label{Eq:alphaMWHH}
\ee

We fitted the combined experimental $H_{\rm c2}(T)$ data in Fig.~\ref{fig:CaPd2Ge2_critical_H}(b) obtained from the $\rho(T)$ and $C_{\rm p}(T)$ measurements by Eqs.~(\ref{Eqs:WHHEqns}) using $\lambda_{\rm so} = 0$, $dH_{\rm c2}(T)/dT|_{T=T_{\rm c}} = - 0.115(15)$~T/K and a corresponding $\alpha_{\rm M} = 0.065$ obtained from Eq.~(\ref{Eq:alphaMWHH}) as shown by solid red curve in Fig.~\ref{fig:CaPd2Ge2_critical_H}(b).  The fit yields $H_{\rm c2}(0) = 134.4$~mT\@. The values $H_{\rm c2}^{\rm Orb}(0) = 0.14(2)$~T in the clean limit and $H_{\rm c2}^{\rm Orb}(0) = 0.13(2)$~T in the dirty limit are estimated from Eq.~(\ref{Eq:Hc2Orb}) using the same value $dH_{\rm c2}(T)/dT|_{T=T_{\rm c}} = - 0.115$~T/K\@.

The Ginzburg-Landau parameter $\kappa_{\rm GL}$ is obtained from $H_{\rm c2}$ and $H_{\rm c}$ using the relation \cite{Tinkham1996}
\be
\kappa_{\rm GL} = \frac{H_{\rm c2}}{\sqrt{2}\,H_{\rm c}},
\label{Eq:KappaGL}
\ee
which for $H_{\rm c2} = 134.4$~mT and the above-obtained $H_{\rm c} = 15.0$~mT gives $\kappa_{\rm GL} = 6.34$. This value of $\kappa_{\rm GL}$ is much larger than $1/\sqrt{2}$ indicating that ${\rm CaPd_2Ge_2}$ is a type-II superconductor.  The lower critical field $H_{\rm c1}$ is obtained from $\kappa_{\rm GL}$ and $H_{\rm c}$ using the relation \cite{Tinkham1996}
\be
H_{\rm c1} = H_{\rm c} \frac{\ln \kappa_{\rm GL}}{\sqrt{2}\,\kappa_{\rm GL}},
\label{Eq:Hc1fromKappaGL}
\ee
yielding $H_{\rm c1}(0) = 3.1$~mT\@.

The characteristic Ginzburg-Landau coherence length at $T=0$, $\xi(0)$, is obtained from \cite{Tinkham1996, DeGennes1966}
\be
H_{\rm c2}(0) = \frac{\Phi_0}{2\pi \xi(0)^2},
\label{Eq:xiFromHc2}
\ee
where $\Phi_0 = 2.07 \times 10^{-7}$~G\,cm$^2$ is the flux quantum. For $H_{\rm c2} = 134.4$~mT we obtain $\xi(0) = 49.7~{\rm nm}$, which is much larger than the mean free path $\ell = 2.55$~nm in Table~\ref{tab:SCParams}, indicating dirty-limit superconductivity in ${\rm CaPd_2Ge_2}$.

The effective magnetic penetration depth $\lambda_{\rm eff}$ is estimated from $H_{\rm c2}$ and $H_{\rm c}$ using the relation \cite{Tinkham1996}
\be
\lambda_{\rm eff} = \frac{\Phi_0 H_{\rm c2}}{4\pi H_{\rm c}^2},
\label{eq:lambda_eff1}
\ee
which gives $\lambda_{\rm eff} = 313$~nm. Alternatively $\lambda_{\rm eff}$ can be estimated using the dirty limit relation \cite{Tinkham1996}
\be
\lambda_{\rm eff}(0) = \lambda_{\rm L}(0)\sqrt{1+\frac{\xi_0}{\ell}}\qquad {\rm (dirty\ limit)}.
\label{eq:lambda_eff2}
\ee
where $\xi_0$ is BCS coherence length which is related to $\xi(T)$ as \cite{Tinkham1996}
\be
\frac{\xi(T)}{\xi_0} =  \frac{\pi}{2\sqrt{3}}\frac{H_{\rm c}(0)}{H_{\rm c}(T)} \frac{\lambda_{\rm L}(0)}{\lambda_{\rm eff}(T)},
\label{eq:xi0l}
\ee
which for $T=0$ leads to
\be
\frac{\xi(0)}{\xi_0} =  \frac{\pi}{2\sqrt{3}} \frac{\lambda_{\rm L}(0)}{\lambda_{\rm eff}(0)}.
\label{eq:xi0l2}
\ee
Substituting $\lambda_{\rm eff}$ from Eq.~(\ref{eq:lambda_eff2}) leads to
\be
\frac{\xi(0)}{\xi_0} = \frac{\pi}{2\sqrt{3\left(1+\frac{\xi_0}{\ell}\right)}}\qquad {\rm (dirty\ limit)}.
\label{Eq:xi0xi00}
\ee
Solving for $\xi_0$ we obtain $\xi_0 = 1181$~nm and then Eq.~(\ref{eq:lambda_eff2}) gives $\lambda_{\rm eff}(0) = 272$~nm in the dirty limit which is similar to the previous estimate of $\lambda_{\rm eff}(0)$.

In order to check the consistency of our parameters we estimate the Fermi velocity from the BCS coherence length which for $\alpha$-model are related by \cite{Johnston2013}
\be
\xi_0 = \frac{\hbar v_{\rm F}}{\pi \Delta(0)} = \left(\frac{1}{\pi \alpha}\right)\frac{\hbar v_{\rm F}}{k_{\rm B} T_{\rm c}}.
\label{eq:xivF}
\ee
We obtain $v_{\rm F} = 1.33 \times 10^8$~cm/s for $\alpha=1.62$ and $\xi_0 = 1181$~nm.  The value of $v_{\rm F}$ is in reasonable agreement with $v_{\rm F}  = 1.53 \times 10^8$~cm/s estimated from Eq.~(\ref{eq:vF}) above from the density of states at the Fermi energy derived from $C_{\rm p}(T)$.

\section{\label{Sec:LondonPD} London Penetration Depth}

\begin{figure}
\includegraphics[width=3.3in]{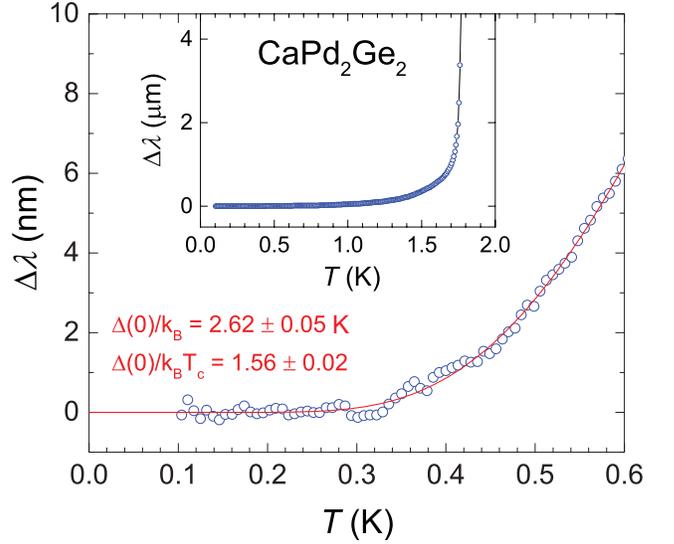}
\caption {The temperature $T$ variation of the $ab$-plane London penetration depth $\Delta\lambda$ in a single crystal of CaPd$_2$Ge$_2$ for the low-$T$ region $0.1~{\rm K} < T < 0.6$~K\@.  The red solid curve is a fit of the data by the BCS prediction in Eq.~(\ref{eq:swavelambda}) for a weak-coupling single-gap $s$-wave superconductor.  The fit parameter is $\Delta(0)/k_{\rm B}$ with the value shown. Inset: $\Delta\lambda$ versus~$T$ over the wide temperature range $0.1~{\rm K} < T < 1.8$~K\@. }
\label{fig:lambda}
\end{figure}

A sample with dimensions of $0.86\times0.61\times0.12$~mm$^3$ with the shortest length being along the tetragonal $c$~axis was cut from the crystal used for the above heat capacity measurements. Figure~\ref{fig:lambda} shows the temperature variation of the $ab$-plane London penetration depth $\Delta\lambda$. The absolute value of the penetration depth was obtained using the TDR technique by matching the frequency shift $\Delta f(T)$ to the skin depth $d(2~\textmd{K})=38.7~\mu$m calculated from the residual resistivity $\rho_0 = 12~\mu\Omega$\,cm and frequency $f=20.2$~MHz.  The superconducting transition temperature was determined as the temperature of the maximum of $d(\Delta\lambda)/ dT$. The $T_c$ obtained is 1.77~K, which lies between the values of 1.69~K and 1.98~K determined from the above heat capacity and resistivity measurements, respectively.

The temperature dependence of $\Delta\lambda$ up to $T_c$ is shown in the inset of Fig.~\ref{fig:lambda}.  At low temperatures below $T_c/3$, the $\Delta\lambda$ in Fig.~\ref{fig:lambda} shows a clear saturation on cooling, which is an indication of a fully-gapped superconducting order parameter in CaPd$_2$Ge$_2$.  CaPd$_2$Ge$_2$ is a dirty-limit superconductor, for which the London penetration depth in the single-band model for $T/T_{\rm c} \ll 1$ is\cite{Anand2013a}
\begin{equation}
\Delta\lambda(T)=\lambda(0)\sqrt{\frac{\pi\Delta(0)}{2k_{\rm B} T}}\exp\left[-\frac{\Delta(0)}{k_{\rm B} T}\right].
\label{eq:swavelambda}
\end{equation}
The experimental data are fitted well up to $T_c/3$ by Eq.~(\ref{eq:swavelambda}) as shown by the solid red curve in Fig.~\ref{fig:lambda}, where the fitting parameters are $\lambda(0)=(186 \pm 16)$~nm and $\Delta(0)/k_{\rm B} = (2.62 \pm 0.05)$~K\@. The error bars are statistical and are obtained from fitting with a fixed temperature domain from $T_{\rm min}$ to $T_{\rm c}/3$.

Using the bulk $T_{c}=1.69$~K determined by the above heat capacity measurement, we obtain $\alpha=1.56(2)$ which is in good agreement with $\alpha=1.62(3)$ determined from the heat capacity jump, both of which are smaller than the value $\alpha_{\rm BCS}\approx 1.764$ expected for an isotropic weak-coupling BCS superconductor. This reduction is likely due to a moderate anisotropy of a single $s$-wave superconducting order parameter\cite{Johnston2013} although the possibility of multigap $s$-wave superconductivity in CaPd$_2$Ge$_2$ cannot be ruled out.\cite{Johnston2013}

The measured and derived superconducting state parameters of ${\rm CaPd_2Ge_2}$ obtained in this and the above sections are summarized in Table~\ref{tab:SCParams}.

\section{\label{Sec:Ani_Gap}Anisotropy in the Superconducting Gap}

The single-band $\alpha$-model of the clean-limit BCS \mbox{$s$-wave} theory of superconductivity$^{22,23}$ describes deviations in the electronic superconducting thermodynamic properties from the BCS predictions by allowing the parameter $\alpha \equiv \Delta(0)/k_{\rm B}T_{\rm c}$ to vary when fitting the superconducting-state electronic heat capacity $C_{\rm e}(T)$, entropy and thermodynamic critical field, whereas the BCS value $\alpha = \alpha_{\rm BCS} \approx 1.764$ is used when calculating $\Delta(T)/k_{\rm B}T_{\rm c}$.  This is an inconsistency in the model, but it does allow the experimental data to be fitted.  Values of $\alpha > \alpha_{\rm BCS}$ indicate strong-coupling superconductivity whereas values $\alpha < \alpha_{\rm BCS}$ indicate anisotropy in the \mbox{$s$-wave} gap around the Fermi surface. \cite{Johnston2013,Kogan2014}  For $\alpha < \alpha_{\rm BCS}$, the $C_{\rm e}(T)$ for $T\lesssim T_{\rm c}$ measures an average over the anisotropic gap, whereas $\lambda(T)$ for $T\ll T_{\rm c}$ measures the minimum value of the gap.

Our $C_{\rm p}(T)$ and $\lambda(T)$ data yield similar values $\alpha = 1.62(3)$ and~1.56(2), respectively.  Thus both measurements indicate the presence of anisotropy in the superconducting $s$-wave order parameter.  However, the BCS theory and the $\alpha$-model based on it were formulated for clean-limit superconductors, whereas we have shown that ${\rm CaPd_2Ge_2}$ is in the dirty limit, which would be expected to reduce the degree of anisotropy that would be present in the clean limit.  Therefore we do not attempt to derive a quantitative estimate of the anisotropy of the superconducting gap in terms of our $\alpha$ values.

\section{\label{Conclusion} Summary}

Single crystals of ${\rm CaPd_2Ge_2}$ were synthesized by the self-flux high-temperature growth technique.  Our powder x-ray diffraction investigation of crushed crystals confirms that the compound crystallizes in the body-centered tetragonal ${\rm ThCr_2Si_2}$- type structure as previously reported.  The superconducting- and normal-state properties were studied using $\rho(T,H)$, $\chi(T)$, $M(H)$, $C_{\rm p}(T,H)$ and $\lambda(T)$  measurements which provide conclusive evidence of bulk superconductivity with $T_{\rm c} = 1.69(3)$~K\@. The normal-state $\rho (T)$ data reveal metallic character and were analyzed using the Bloch-Gr\"{u}neisen model of electrical resistivity. The normal-state $C_{\rm p}(T)$ data were analyzed using the sum of an electronic term and the Debye model of lattice heat capacity, yielding the Sommerfeld coefficient $\gamma_{\rm n} = 8.78(4) $~mJ/mol\,K$^{2}$ and Debye temperature $\Theta_{\rm D} = 259(2)$~K\@. The normal-state $\chi(T)$ and $M(H,T)$ data exhibit weak anisotropic $T$-independent paramagnetism with $\chi_{ab} > \chi_{c}$.

The superconducting transition is revealed by a very sharp jump in $C_{\rm p}$ at $T_{\rm c} = 1.69(3)$~K with a heat capacity jump at $T_{\rm c}$ given by $\Delta C_{\rm e}/\gamma_{\rm n} T_{\rm c}  =1.21(3)$ that is smaller than the BCS value of 1.426. The superconducting state electronic heat capacity data are analyzed using the $\alpha$-model of the BCS theory of superconductivity. The very good agreement between the $\alpha$-model prediction and the experimental data supports the applicability of the model. The value $\alpha \equiv \Delta(0)/k_{\rm B}T_{\rm c} = 1.62(3)$ obtained is smaller than the BCS value $\alpha_{\rm BCS} = 1.764$, reflecting the corresponding reduction in the heat capacity jump at $T_{\rm c}$. Within a single band model, this reduction arises from anisotropy in an $s$-wave gap, but multigap $s$-wave superconductivity is not ruled out.  Various normal- and superconducting-state parameters are estimated which indicate that ${\rm CaPd_2Ge_2}$ is a dirty-limit weak-coupling type-II $s$-wave electron-phonon driven superconductor.

\acknowledgments

We thank V. G. Kogan for helpful comments on the gap anisotropy. This research was supported by the U.S. Department of Energy, Office of Basic Energy Sciences, Division of Materials Sciences and Engineering.  Ames Laboratory is operated for the U.S. Department of Energy by Iowa State University under Contract No.~DE-AC02-07CH11358.


\end{document}